\documentclass[12pt,preprint]{aastex}

\begin{document}

\title{Neutrinos From Individual Gamma-Ray Bursts in the BATSE Catalog}

\author{D. Guetta\altaffilmark{1}, D. Hooper\altaffilmark{2},
J. Alvarez-Mu\~niz\altaffilmark{3}, F. Halzen\altaffilmark{2},
E. Reuveni\altaffilmark{4} }

\altaffiltext{1}{Osservatorio astrofisico di Arcetri, L.E. Fermi 5, 50125 
Firenze, Italy; dafne@arcetri.astro.it}
\altaffiltext{2}{Department of Physics, University of Wisconsin, 
1150 University Avenue, Madison, WI  53703; hooper@pheno.physics.wisc.edu;
halzen@pheno.physics.wisc.edu}
\altaffiltext{3}{Departamento de F\'\i sica de Part\'\i culas, Facultade de F\'\i sica, 15706 Santiago de Compostela, A Coru\~na, Spain; jaime@fpaxp2.usc.es}
\altaffiltext{4}{The Hebrew University of Jerusalem, 
Faculty of Agriculture, Rehovot, Israel; ftdeli@agri3.huji.ac.il}
\begin{abstract}

We estimate the neutrino emission from individual gamma-ray bursts 
observed by the BATSE detector on the Compton Gamma-Ray Observatory. 
Neutrinos are produced by photoproduction of pions when protons 
interact with photons in the region where the kinetic energy 
of the relativistic fireball is dissipated allowing the acceleration
of electrons and protons.
We also consider models where neutrinos are predominantly produced on 
the radiation surrounding the newly formed black hole. From 
the observed redshift and photon flux of each individual burst, we 
compute the neutrino flux in a variety of models based on the 
assumption that equal kinetic energy is dissipated into electrons and 
protons. Where not measured, the redshift is estimated by other 
methods. Unlike previous calculations of the universal diffuse neutrino 
flux produced by all gamma-ray bursts, the individual fluxes (compiled 
at http://www.arcetri.astro.it/$\sim$ dafne/grb/) can be directly compared 
with coincident observations by the AMANDA telescope at the South Pole.
Because of 
its large statistics, our predictions are likely to be representative 
for future observations with larger neutrino telescopes.

\end{abstract}

\keywords{gamma rays: bursts---acceleration of particles---neutrinos}

\section{Introduction}
\label{sec:intro}

The leading models for Gamma-Ray Bursts (GRB), bursts of 0.1-1 MeV
photons typically lasting for 0.1-100 seconds (Fishman \& Meegan 1995),  
involve a relativistic wind emanating from a compact central source. 
The ultimate energy source is rapid accretion onto a newly formed stellar 
mass black hole. 

Observations suggest that the prompt $\gamma$-ray emission is 
produced by the dissipation (perhaps due to internal shocks)
of the kinetic energy of a relativistically expanding wind, 
i.e. a ``fireball''.
Both synchrotron and inverse Compton emissions from the shock-accelerated 
electrons have been proposed as the GRB emission mechanism.

In this paper, we study in detail the production of neutrinos by 
protons accelerated along with electrons. We assume that equal energy of the fireball is dissipated in protons and electrons (or photons). This is the case in models where GRBs are the sources of the highest energy cosmic rays. The basic idea is that the 
protons produce pions decaying into neutrinos in interactions with the fireball photons, or with external photons surrounding the newly formed black hole. 

Where previous calculations have estimated the universal diffuse flux of neutrinos produced by all GRBs over cosmological time, we estimate the flux from individual GRBs observed by the BATSE (Burst And Transient Source Experiment) experiment on the Compton Gamma-Ray observatory. 
The prediction can be directly compared with coincident observations performed with the AMANDA detector. Having these observations in mind, we specialize on neutrino emission coincident in time with the GRB. Opportunities for neutrino production exist after and, in some models, before the burst of $\gamma$-rays, e.g. when the fireball expands through the opaque ejecta of a supernova.

The calculations are performed in two models chosen to be representative and rather different versions of a large range of competing models. The first is generic for models where an initial event, such as a merger of compact objects or the instant collapse of a massive star to a black hole, produces the fireball (Waxman \& Bahcall 1997 ;Guetta, Spada \& Waxman 2001a, GSW hereafter). 
We calculate the 
neutrino production by photomeson interactions of relativistic
protons accelerated in the internal shocks and the synchrotron 
photons that are emitted in these shocks. 
The neutrinos produced via this mechanism have typical energies of 
$\sim 10^{14}-10^{15}$ eV, and are emitted in coincidence with the GRBs,
with their spectrum tracing the GRB photon spectrum.
For an alternative model, we have chosen the supranova model where a 
massive star collapses to a neutron star with mass $\sim 2.5-3\,M_\odot$, 
which loses its rotational energy on a time scale $t_{\rm sd}$ of weeks to 
years, before collapsing to a black hole, thus triggering the GRB. 
Following Guetta and Granot (Guetta \& Granot 2002a), we calculate the 
neutrino flux from interactions of the fireball protons with external photons in the rich radiation field created during the spindown of the supra-massive pulsar. Production on external photons turns out to be dominant for a wide range 
of parameters, $t_{\rm sd}\lesssim 0.2\;$yr for a typical
GRB and $t_{\rm sd}\lesssim 2\;$yr for X-ray flashes.
The neutrinos produced via this mechanism have energies 
$\varepsilon_\nu\sim 10^{15}-10^{17}\,(10^{19})\;$eV for typical GRB 
(X-ray flashes) and are emitted 
simultaneously with the prompt $\gamma$-ray (X-ray) emission.
Their energy spectrum consists of several power law segments and its overall shape depends on the model parameters, especially $t_{\rm sd}$. If the mass of the supernova remnant is of the order of $\sim 10 M_{\odot}$,
and if the supernova remnant shell is clumpy, then 
for time separation $t_{\rm sd}<0.1$ yr  the SNR shell has a Thomson optical
depth larger than unity and obscures the radiation 
emitted by the GRB. 
Therefore, for $t_{\rm sd}\lesssim 0.1\;$yr, the $\nu$'s would not 
be accompanied by a detectable GRB providing us with an example of neutrino emission not coinciding with a GRB display.

As previously mentioned, to realistically estimate the neutrino fluxes associated with either of these models, we turn to the GRB data collected by BATSE. The BATSE records include spectral and temporal information which can be used to estimate neutrino 
spectra for individual bursts. We will perform these calculations for two version of each of the two models previously described. The two versions correspond to alternative choices of important parameters. The wide variety of GRB spectra results, not surprisingly, in a wide range of neutrino spectra and event rates. For approximately 800 bursts in the BATSE catalog, and for four choices of models, we have calculated the neutrino spectra and the event rates, coincident with GRBs, for a generic neutrino telescope. With 800 bursts, the sample should also be representative for data expected from much larger next-generation neutrino observatories. 

Neutrino telescopes can leverage the directional and time information provided by BATSE to do an essentially background-free search for neutrinos from GRBs. Individual neutrino events within the BATSE time and angular window are a meaningful observation. 
A generic detector with 1\,km$^2$ effective telescope area, during one year, should be able to observe 1000 bursts over $4\pi$ steradians. 
Using the BATSE GRBs as a template, we predict order 10 events, 
muons or showers, for both models. 
The rates in the supranova model depend
strongly on $t_{\rm sd}$. 
In this model we anticipate $\sim 7$ events per year 
assuming $t_{\rm sd}\simeq 0.07$ yr, but only one event per ten years 
for $t_{\rm{sd}}\simeq 0.4$ yr. 
We will present detailed tabulated predictions 
further on. They can be accessed at 
http://www.arcetri.astro.it/$\sim$dafne/grb/.

Short duration GRBs, characterized by lower average fluences, are less likely to produce observable neutrino fluxes.

We find that GRBs with lower peak energies, X-ray flash candidates, yield the largest rates in the supranova model. For instance, for $t_{\rm{sd}}\simeq 0.07$ yr, we predict one event (muon or shower) for every 1000 bursts. If only 100, or so, X-ray flashes occur per year, as observations suggest, this will be difficult to observe.  However, such events may be considerably more common and may contribute significantly to the diffuse high energy neutrino flux. Observations of neutrinos from this class of GRBs would be strong evidence for a supranova progenitor model.

The AMANDA collaboration has collected neutrino data in coincidence with BATSE observations (Barouch \& Hardtke 2001). It operated the detector for 3 years (1997-1999) with an effective area of approximately 5,000\,m$^2$ for GRBs events. In early 2000 the
 expanded detector reached an effective area of roughly 50,000\,m$^2$. Unfortunately, only $\sim$\,100 coincident bursts could be observed with the completed detector before BATSE operations ceased in June 2000. With effective areas significantly below the canonical square kilometer discussed in this paper, AMANDA is not large enough to test the GRB models considered here. We estimate only 0.08 events in 1997-1999 and 0.3 in 2000.

Our estimate of 1--10 events in 1 year for a telescope with 1
kilometer square telescope area is consistent with previous,
burst-averaged, determinations of the GRB neutrino flux (Alvarez, Halzen 
\& Hooper 2000, Dermer \& Atoyan 2003). Note that the effective area 
of IceCube for
GRB will significantly exceed this reference value (PDD). The rate can
be understood by a back-of-the-envelope estimate. A typical GRB
produces a photon fluence on the order of $10^{-5}\,
\rm{ergs}/\rm{cm}^2$, which we assume to be equal to the energy in
protons. If $20\%$ of the proton energy is converted into pions, half
to charged pions, and one quarter of the charged pion energy to muon
neutrinos in the $\pi \rightarrow \mu \rightarrow e$ decay chain, then
$\sim 5 \times 10^{-7}\, \rm{ergs}/\rm{cm}^2 /E_{\nu}$ neutrinos are
generated. For a typical neutrino energy of $E_{\nu}\sim
100 \, \rm{TeV}$, this yields $\sim 30$ neutrinos per square
kilometer. At this energy the probability that a neutrino converts to
a muon within range of the detector is $10^{-4}$ (Gaisser, Halzen,
Stanev 1995). Therefore $3 \times 10^{-3}\,$ muons are detected in
association with a  single GRB. With over 1000 GRBs in one year, we
estimate a few muons per year in a
kilometer-square detector. Fluctuations in fluence and other burst
characteristics enhance this estimate significantly (see Figure \ref{figevn},
for instance), however, absorption
of neutrinos in the Earth can reduce this number. These effects are
included in the calculations of actual event rates throughout this
paper.

Our estimates, while observable with future kilometer-scale observatories, may be conservative. We already mentioned bursts with no counterpart in photons. Also, occasional nearby bursts, much like supernova, could exceptionally provide higher event rates than our calculations reflect. 
We would like to point out the fact that the aim of the paper is not
to do more precise calculations than the ones already done in the literature.
We just want to generate results that can be compared with 
experiments that do coincidents observations with satellites.

The outline of the paper is somewhat unconventional. The detailed results are collected in a data archive at http://www.arcetri.astro.it/$\sim$dafne/grb. The details of the calculations of the neutrino fluxes in the two models are described in appendices 
A and B. In appendix C we collected the methods used to evaluate the rates of neutrino-induced muons, taus and showers in a generic detector with a given effective telescope area. The main body of the paper is organized as follows. 
In \S 2, we review
 GRB models and describe the mechanisms for neutrino production in GRBs. 
In \S 3 and \S 4 we describe the BATSE catalog of GRBs and we 
identify subclasses of events: long duration GRBs with and without  measured redshift, short duration GRBs and X-ray 
flash candidates. In \S 5 we summarize our simulation of the response of a generic high energy neutrino telescope to the predicted neutrino fluxes. 
In \S 6 we analyse some anomalous bursts. 
Results and conclusions are collected in \S 7.

\section{GRB Models and Neutrino Emission}
\label{NuEm}

Progenitor models of GRBs are divided into two main categories. 
The first category involves the merger of a binary system of compact 
objects, such as a double neutron star (Eichler et al. 
1989), a neutron star and a black hole (Narayan, 
Pacy\'nski \& Piran 1992) or a black hole and a Helium star or a white dwarf 
(Fryer \& Woosley 1998; Fryer, Woosley \& 
Hartmann 1999). The second category involves the death of a massive star. 
It includes the failed supernova (Woosley 1993) or hypernova 
(Pacy\'nski 1998) models, where a black hole is created 
promptly, and a large accretion rate from a surrounding accretion disk 
(or torus) feeds a strong relativistic jet in the polar regions. 
This type of model is known as the collapsar model.
An alternative model within this second category is the supranova model 
(Vietri \& Stella 1998),
where a massive star explodes in a supernova and leaves behind a 
supra-massive neutron star (SMNS), of mass $\sim 2.5-3\,M_\odot$. 
It subsequently loses its rotational energy on a time scale $t_{\rm sd}$ of order weeks to years until it  collapses to a black hole. This triggers the GRB. 
Long GRBs (with a duration $\gtrsim 2\;{\rm s}$) are usually attributed 
to the second category of progenitors, while short GRBs are attributed to 
the first category. 

We select two models to investigate the opportunities for neutrino production in a GRB. The analysis can be easily extended to other models
(see for example Razzaque, M\'esz\'aros \& Waxman 2002; Dermer \& Atoyan 2003).
The first model is based on the standard fireball phenomenology where electrons and protons are shock accelerated in the fireball. Pions and neutrinos are produced by photoproduction interactions when the protons coexist in the fireball with photons. These are produced by synchrotron radiation of accelerated electrons. For a second model we turn to the supranova scenario where the 
supra-massive pulsar loses its rotational 
energy through a strong pulsar wind. This pulsar wind creates a rich external radiation field before the collapse to the final black hole and the creation of the GRB fireball. It is referred to as the pulsar wind bubble (PWB) (K\"onigl \& Granot 2002;  
Inoue, Guetta \& Pacini 2002; Guetta \& Granot 2003) 
and provides a target for the photoproduction of neutrinos by fireball protons. Let's note in passing that the supranova model has several advantages compared to other collapsar 
models: (i) the jet does not have to penetrate the stellar 
envelope (Vietri \& Stella 1998), 
(ii) it can naturally explain the X-ray line features observed in several 
afterglows (Piro et al. 2000; Lazzati, et al. 2001;
Vietri et al. 2001), and
(iii) the large fraction of the internal 
energy in the magnetic field and in electrons observed in the afterglow emission arise 
naturally (K\"onigl \& Granot 2002; Guetta \& Granot 2003).

In all of the different scenarios mentioned above, the final stage of the 
process consists of a newly formed black hole with a large accretion rate 
from a surrounding torus, and involve a similar 
energy budget ($\lesssim 10^{54}\;{\rm ergs}$). Observations suggest that prompt $\gamma$-ray emission is 
produced by the dissipation of the kinetic energy 
within the fireball, due to internal shocks within 
the flow that arise from variability of the Lorentz factor, $\Gamma$,
on a time scale $t_v$. The afterglow emission is thought to arise from an external shock 
that is driven into the ambient medium as it decelerates the ejected matter 
(Rees \& M\'esz\'aros 1994; Sari \& Piran 1997).
In this so called `internal-external' shock model, 
the duration of the prompt GRB is directly related to the time during 
which the central source is active. 
The emission mechanism is successfully described by synchrotron radiation
from relativistic electrons that radiate 
in the strong magnetic fields. These are close to equipartition values within the shocked plasma. 
An additional radiation mechanism that may also play some role is 
synchrotron self-Compton (SSC) (Guetta \& Granot 2002b), 
which is the upscattering of the synchrotron photons by relativistic 
electrons to higher energy. 

Protons are expected to be accelerated along with the electrons in the
region where the wind kinetic energy is converted into internal energy
due to a dissipation mechanism like internal shocks.
The conditions in the dissipation region allow proton 
acceleration up to $\varepsilon_{p,{\rm max}}\gtrsim 10^{20}\;$eV 
(Waxman 1995; Vietri 1995). The energy in $\gamma$-rays reflect the fireball energy in accelerated electrons and afterglow observations
indicate that accelerated electrons and protons carry similar energy (Freedman \& Waxman 2000). Our basic assumption in calculating neutrino emission from GRBs is that equal amounts of energy go into protons and photons. In models where GRB protons are the source of the highest energy cosmic rays, this assumption is supported by the approximate equality  of the $\gamma$-ray fluence of all GRBs and the total energy in extragalactic cosmic rays.

Both internal shocks, responsible for the prompt GRB emission,
and the external shock, responsible for the afterglow emission,
have been proposed as possible sources of the highest energy cosmic rays (Waxman 1995 and Vietri 1995, respectively). A comparison between the two mechanisms has been done by
Vietri, De Marco and Guetta (2003), but it is not easy to conceive, at this point, an observational
test capable of distinguishing between the two models. The only hope appears to observe the production of high energy neutrinos which must accompany the {\it in situ} acceleration
of particles. Occasionally, ultra-high energy protons will produce pions and neutrinos in collisions with photons in photon rich environment provided by the post--shock shells or by the pulsar wind bubble.
If the protons are accelerated in internal shocks, the neutrinos produced will arrive at Earth simultaneously with the photons
of the burst proper and will have an energy $\sim 10^{15}-10^{16}\, \rm{eV}$
(Waxman and Bachall 1997; GSW; Guetta \& Granot 2002a).
If accelerated in external
shocks, they will arrive at Earth simultaneously with the photons
of the afterglow and will have a higher energy, $\gtrsim 10^{17}$ eV
(Vietri 1998a, 1998b, Waxman \& Bahcall 2000).

General phenomenological considerations indicate that gamma-ray 
bursts are produced by the dissipation of the kinetic energy 
of a relativistic expanding fireball. Internal shocks that are mildly
relativistic are believed to dissipate the energy. Therefore, the proton energy distribution
should be close to that for Fermi acceleration
in a Newtonian (non-relativistic) shock, $dn_p/d\varepsilon_p\propto\varepsilon_p^{-2}$.
Moreover, the power law index of the electron and proton energy distributions
are expected to be the same, and the values inferred for the electron distribution from the observed photon spectrum are $dn_e/d\varepsilon_e\propto\varepsilon_e^{-p}$
with $p\sim 2-2.5$. We shall, therefore, adopt
$dn_p/d\varepsilon_p\propto\varepsilon_p^{-2}$.
Plasma parameters in the dissipation region allow proton acceleration
to energies greater than $10^{20}$ eV (Waxman 1995, Vietri 1995).

We will assume that the fireball is 
spherically symmetric. Note, however, that a jet-like  fireball behaves as if it were a conical section of a
spherical fireball as long as $\theta_j>1/\Gamma$, where $\theta_j$ is the
jet opening angle and $\Gamma\gtrsim 300$ is the wind Lorentz factor.
Therefore, our results apply without modification to a jet-like fireball.
For a jet-like wind, the luminosity, $L$, in our equations should be understood as the
luminosity of the fireball inferred by assuming spherical symmetry.

We have relegated all details of the calculation of neutrino production via photomeson interaction with GRB photons and PWB photons to appendices A and B, respectively. Throughout the paper, we will refer to the models by the following convention:

\begin{itemize}

\item[(i)] Model 1: Neutrino flux from the interaction of high-energy protons with GRB photons.  The fraction of proton energy transfered to pion energy is set to 0.2 as indicated by simulations of GSW (see apppendix A).

\item[(ii)] Model 2: Neutrino flux from the interaction of high-energy protons with GRB photons.  The fraction of proton energy transfered to pion energy is calculated as described in appendix A (see appendix A).

\item[(iii)] Model 3: Neutrino flux from the interaction of high-energy protons with PWB photons, as in the supranova progenitor model. The time scale between supernova and GRB is set to $t_{\rm{sd}}=0.4$ yr (see appendix B).

\item[(iv)] Model 4: Neutrino flux from the interaction of high-energy protons with PWB photons, as in the supranova progenitor model. The time scale between supernova and GRB is set to $t_{\rm{sd}}=0.07$ yr (see appendix B).

\end{itemize}

\section{The BATSE Catalog}

BATSE, the Burst And Transient Source Experiment, was a high energy 
astrophysics experiment launched on the Compton Gamma-Ray Observatory in 1991.
BATSE, between its launch and the termination of its orbit in 2000, 
has observed and recorded data from over 8000 events including gamma-ray 
bursts, pulsars, terrestrial gamma-ray flashes, soft gamma repeaters and 
black holes.

The data BATSE recorded from gamma-ray bursts is publically available in 
the current BATSE catalog at http://f64.nsstc.nasa.gov/batse/grb/catalog/. 
For a description see Paciesas et al. 1999. 
The catalog includes information on the spectrum, time and location 
of each triggered burst. Each triggered event has been assigned a 
BATSE trigger number (between 105 and 8121 for GRBs) which we use to 
identify individual bursts.

Spectral information is recorded in four energy channels, 
20-50 keV, 50-100 keV, 100-300 keV and above 300 keV. 
Using these four fluence measurements, we have fitted the spectrum of each GRB 
to a broken power law, treating the break energy, both spectral slopes and 
the normalization as free parameters; see Eq.(\ref{eq:Fnu}). 
We determine the Lorentz factor of the relativistic expanding ejecta 
using the break energy through Eq.(\ref{eq:epeak}).
For bursts 
with an observed break energy above 300 keV, or below 50 keV, it is difficult to determine both the break energy and the power law slope. For high energy breaks, the impact on this ambiguity is not critical. As explained in appendix A, the Lorentz factor 
is not dependent on the fit because it is fixed by the 
observed high energy of the event and the requirement that the fireball be
optically thin. 
For very low spectral breaks (below 50 keV), for instance in 
events we classify as X-ray flash candidates, we acknowledge a significant 
degree of uncertainty in the Lorentz factor calculation and resulting 
neutrino spectra. In this case, the spectral break is only uncertain 
to about a factor of 2 or 3.

Detailed temporal information is available in the BATSE catalog, 
in the form of light curves.  The BATSE time resolution varies between 
2.048 seconds and 0.016 seconds. A resolution of 0.064 seconds is 
available for all bursts during the time following the trigger. 

In the framework of the internal shock model,
we need a variability time, $t_v$,  $\lesssim 0.01 s$ in order 
to get the 1MeV $\gamma-rays$ (Guetta, Spada \& Waxman 2001b, Waxman 2001). 
In fact with a value of the Lorentz factor larger than the minimal
value needed to be optically thin up to 100 Mev  ($\sim 250$ see 
Eq.\ref{eq:gamma}) 
the variability time,  $t_v$, has to be $\lesssim 10 ms$ to get the 1 MeV 
$\gamma$-rays.

How well the data support the model is still controversial and a detailed
analysis on this issue is out of the aim of our paper.
Since we refer to this model for our analysis we consider a value of
$t_v=10$ ms for long duration GRBs in the rest of the paper. 
For short duration bursts we take $t_v=0.1$ ms 
and 0.050 seconds for X-ray flash candidates.

\section{GRB Classes}
We have divided the list of BATSE GRBs into four different classes: 
1) long duration bursts (duration $\gtrsim 2\;{\rm s}$) with measured redshfit, 
2) long duration bursts without measured redshift, 
3) short duration bursts and 4) X-ray flashes. 
Major differences in temporal and spectral properties of long duration GRBs, 
short duration GRBs and X-ray flashes has lead to some speculation that 
they may involve different progenitors or mechanisms.

\subsection{Long Duration GRBs With Measured Redshift}
\label{zknown}

By observing the optical afterglow of GRBs, it is possible to measure 
spectral lines and, therefore, the redshift of an individual burst. 
Although to date the X-ray afterglow of on the order of 
100 long duration GRBs 
have been observed, only 31 of them have been observed in the optical making a determination of the redshift possible. (No redshift has been identified for short GRBs). We first consider 13 of these that have a 
complete BATSE record.

The relationship between the comoving distance to an object and 
its redshift is given by: 
\begin{eqnarray}
d =\frac{c}{H} \int^z_0 \frac{dz' }{  
\sqrt{  \Omega_{\Lambda}+\Omega_{\rm{M}}(1+z' )^3  }}
\end{eqnarray}
Where $\Omega_{\Lambda}\simeq 0.7$ and $\Omega_{\rm{M}}\simeq 0.3$ are 
the fractions of the critical density of the Universe in dark energy and 
matter, respectively. $H$ is Hubble's constant.

Once the distance to a GRB is known, and its fluence (or flux) has been 
measured, the isotropic-equivalent $\gamma$-ray luminosity can be 
calculated. From this and the value of $t_v=0.01\; \rm{s}$, together
with the break energy of the GRB photon spectrum, we estimate 
the bulk Lorentz factor using Eqs.(\ref{eq:gamma}, \ref{eq:epeak}).  Setting the efficiency for pion production $f_{\pi}=0.2$ for model 1 and using Eq.(\ref{eq:fpi2}) 
for model 2, we estimate the fraction of proton energy transfered into pions. Using Eqs.(\ref{eq:synclos}, \ref{eq:synclos2}) the energy scale of synchrotron losses, $\varepsilon^{\rm{s}}_{\nu}$, is determined. From these informations we can determine the
 neutrino flux at Earth for each of the four models. For a more detailed discussion see appendices A and B. 

\subsection{Long Duration GRBs Without Measured Redshift}
\label{zunknown}

For the majority of the long duration bursts in the BATSE catalog no redshift is available. In this case, the $\gamma$-ray luminosity of long 
duration bursts, and, therefore, the distance, 
can be estimated by assuming a relationship between the observed variability and the luminosity of a GRB (Lloyd-Ronning \& Ramierz-Ruiz 2002; Zhang \& Meszaros 2002; Kobayashi, Ryde \& MacFadyen 2002). Note that the variability of a GRB is not 
the same as its variability time, $t_v$, previously introduced.

The variability of burst is a measure of fluctuations in the 
temporal structure of the burst.  It is defined such that pure noise should have a variability of zero, while the most variable bursts have very 
sudden and distinctive temporal features.  We use the following 
definition of variability (Fenimore \& Ramirez-Ruiz 2000):
\begin{eqnarray}
V=Y^{-0.24} \frac{1}{N} \Sigma \frac{(F_i - F_{{0.30}
\times {T_{90}}})^2 - (F_i+B)          }{F^2_{\rm{peak}}}
\end{eqnarray}
where $Y=(1+z)/3$, $F_i$ is the background subtracted fluence in a 
time bin $i$, $B$ is the average background in a single time bin, 
$F_{\rm{peak}}$ is the maximum fluence and 
$F_{{.30}\times {T_{90}}}$ is the average fluence over a time period 
centered at time bin $i$ of length 30\% of the $T_{90}$ 
time (duration) of the burst.

The GRBs with observed redshifts have been used to empirically derive a relationship between the variability and the 
luminosity of the burst (Fenimore \& Ramirez-Ruiz 2000; Reichart et al. 2001, 
Reichart \& Lamb 2001):
\begin{eqnarray}
L/ d\Omega = 3.1 \times 10^{56} V^{3.35} \rm{erg}\,\, \rm{s}^{-1}.
\end{eqnarray}
A relationship between luminosity and the time lag between the 
peaks for light curves in different energy bands, has been 
observed in the GRB redshift data (Norris, Marani \& Bonnell 2000), 
but appears to be less reliable.  Therefore, we will only 
consider the luminosity-variability relationship.

Together with the fluence (flux) of a burst, the distance of a 
burst can be determined by the luminosity.  Note that the variability, 
and therefore luminosity, of a burst depends on the redshift or distance 
to the burst.  Therefore, we must do this calculation by iteration.  
The end result is a value of the luminosity and redshift for each burst.
We would like to enphasize the fact that there are a lot of uncertainties 
in this way to estimate the redshift, however the knoweledge of 
the redshift is 
not so important in our analysis since the minimum variability time scales 
and the fluence are the most important quantities.

It is difficult to reliably calculate the variability of bursts with low flux. For this reason, 
we only consider bursts with a peak flux greater than 1.5 photons/$\rm{cm}^2 
\rm{sec}$ (over a 0.256 s time scale) and with at least 30 time bins 
(of 0.064 s width) of 5 sigma or more above the average background. 
The GRBs which do not meet these requirements have low fluence and are 
therefore likely to yield a low neutrino flux anyway. 
After these criteria were applied, 566 long duration bursts without measured 
redshift are left, making it our largest class. 

Even for bursts which meet the above criteria, 
the luminosity estimated is only accurate to an order of magnitude. 
This corresponds to uncertainties of a factor of 2 or 3 in the fraction 
of proton energy transfered into pions and in the synchrotron loss energy.

\subsection{Short Duration GRBs}
\label{short}

Short duration GRBs, with no observation of an optical afterglow and, therefore, no measurement of redshift, cannot have a relationship between variability and luminosity empirically established. Additionally, variability is difficult to measure for short
 duration bursts. Left with no way to measure the $\gamma$-ray luminosity of, or distance to, 
a short duration GRBs, we choose to set $z=1$ for each burst. This introduces greater uncertainty 
than in the long duration GRBs case but, given the ambiguities in the burst 
characteristics, it is the best that can be done at this time. 
To be consistent, we considered only short bursts with a peak flux greater 
than 1.5 photons/$\rm{cm}^2 \rm{sec}$ (over a 0.256 s time scale), as we 
did with for long duration GRBs. After this criteria was applied, 
199 short duration bursts remained in this category.

Temporal structure and variations appear to occur on shorter time scales for short compared to long GRBs (McBreen et al. 2002). We therefore use a time scale of temporal fluctuations of $t_v=0.001 \, \rm{s}$ for all short duration bursts, as opposed to the value of $t_v=0.01 \, \rm{s}$ 
used for long duration bursts. It is also interesting to note that short duration bursts generally 
have a somewhat harder spectrum and higher peak energy than 
long duration GRBs (Paciesas et al. 2001). 

\subsection{X-Ray Flash Candidates}
\label{Xray}

X-ray flashes are a newly discovered class of fast transient sources. 
The BeppoSAX experiment's wide field cameras have observed such events at 
a rate of about four per year (Heise et al. 2001, Kippen et al. 2002), 
implying a total rate on the order of $100$ per year. 
These events typically have peak 
energies as low as 2-10 keV and durations of 10-100 seconds. 
The BeppoSAX experiment discriminates X-ray flashes from standard GRBs by 
the non-detection of a signal above 40 keV with the BeppoSAX GRB monitor. 
More generally, a large ratio of X-ray to $\gamma$-ray fluence is the 
differentiating characteristic of X-ray flashes from GRBs. 
It has been suggested, however, that X-ray flashes are a low 
peak energy extension of gamma-ray bursts 
(Heise et al. 2001, Kippen et al. 2002).

The final class considered here consists of BATSE events which may be X-ray flashes. We identify 15 events in this class with spectra  that peak below 50 keV, although it is difficult to determine accurately 
where the peak occurs because the sensitivity of BATSE is somewhat poor in this energy range. These are long duration bursts, and typically have a hard 
spectrum; several have $\beta$ in Eq.(\ref{eq:Fnu}) larger than 1.5, and no 
observed flux in BATSE's fourth energy channel (above 300 keV).

Again, with no measured redshift, and limited temporal information, 
we cannot deduce the luminosities of these events. We set $z=1$ for each event 
and calculate its luminosity accordingly. It is thought that the radius of collisions in X-ray flashes is typically 
larger than in other GRBs and, therefore, the time scale of fluctuations 
will generally be larger (Guetta, Spada \& Waxman 2001b). 
We therefore choose $t_v=0.05 \,\rm{s}$. For some X-ray flashes the peak energy is very low (20-30 keV) and the Lorentz factor accordingly very high. Increasing the time scale of fluctuations has the 
additional effect of lowering the Lorentz factor to a reasonable value 
in these extreme cases (see Eq.(\ref{eq:epeak})).

For X-ray flashes, with a very large Lorentz factor and a long time scale of 
fluctuations, we expect that a very small fraction of proton energy 
converted to pions; see Eqs.(\ref{eq:fpi1},\ref{eq:fpi2}). We therefore expect low neutrino fluxes from proton interactions with GRB photons. For the models involving proton interactions with photons in a 
surrounding pulsar wind bubble, however, the rates can be quite high 
(Guetta \& Ganot 2002a).

There is another class of objects that are the non-triggered bursts.
They are found in the BATSE data in off-line analysis; see e.g. Stern and
Tikhomirova \\
(http://www.astro.su.se/groups/head/grb$_{-}$archive.html). They
are not energetic enough to trigger in real time.
These can be GRBs with low kinetic luminosity and will be very weak neutrino
sources for all the four models, therefore we have decided to neglect
them in our analysis.

\subsection{Neutrino GRB Database}

We have constructed a database publicly accessible online with a
complete list of all of the GRB characteristics and associated neutrino event rates for the approximately 800 bursts we have considered in this
analysis. The data was populated in a MySQL database, interacting through a user-friendly web interface developed with PERL.  The
database is searchable by GRB number and/or by GRB class, and is
capable of searching for only bursts included in the AMANDA analysis
or all bursts in this work. The rates are for a generic neutrino telescope provided the threshold is sufficiently low for observing the neutrino fluxes predicted. How we transform the GRB neutrino fluxes into observed event rates is the topic of the next 
section. Representative results are tabulated in tables 1 through 12. The database also contains the predicted neutrino spectra for each burst; these can be directly combined with the simulation of a specific detector. The database is accessible at http://www.arcetri.astro.it/$\sim$dafne/grb.

\section{Neutrino Telescopes and Event Simulation}

Large volume neutrino telescopes are required to observe and measure the neutrino flux from GRBs. Current experiments, such as AMANDA (Andres et al. 2001) at the South Pole, or future (Aslanides et al. 1999) and next generation experiments, including IceCube (see http://icecube.wisc.edu/) with a full cubic kilometer of instrumented detector volume, i.e. with $> 1 {\rm km^2}$ telescope area, are designed to observe high energy cosmic neutrinos with energies expected from GRBs.

Neutrino telescopes detect the Cherenkov light radiated by showers (hadronic and electromagnetic), muons, and taus  that are produced in the interactions of neutrinos inside or near the detector. Muons are of particular interest because, at the energies typical for GRB neutrinos, they travel kilometers before losing energy.  The
dominant signal is, therefore, through-going muons.  IceCube can measure
the energy and direction of any observed muon. The angular resolution
is less than $1^\circ-3^\circ$ while the energy resolution is
approximately a factor of three. Signal and background muons may,
therefore, be differentiated with a simple energy cut. For shower events the energy measurement improves significantly, being better than 20\%, but reconstructing their direction is challenging, the angular resolution being
of order 10 degrees.

For high energies, when tau decay is sufficiently time dilated, taus
have a range similar to or larger than muons, and so the dominant tau
signal is from through-going taus.  These events have a characteristic
signature consisting of a ``clean'' minimum-ionizing track despite its long range inside the detector.  We will therefore consider all events in which a tau track passes through the detector in the direction and at the time of a GRB. We assume that taus and muons are distinguishable at all energies when specializing to rare events in coincidence in time and direction with a GRB. 
We realize that, in general, it may be difficult to distinguish a muon of
energy $\lesssim 200$ GeV, that is expected to lose relatively little energy from catastrophic processes, from a very high energy tau.
Tau signatures however become dramatic when they decay inside the detector (lollipop events), or when the tau neutrino interacts and the produced tau decays into showers inside the detector (double bang events). 

To evaluate the prospects for GRB neutrino observations, it is essential to
determine the rate of muon, shower and tau events. These calculations are each described in appendix C. 
For a review of high energy neutrino astronomy; see (Halzen \& Hooper 2002; Learned \& Mannheim 2000).

\section{Anomalous Bursts}

There are a few GRBs we have considered in this analysis which are anomalous for a variety of reasons.  We briefly mention these in this section.

\begin{itemize}

\item[(i)] GRB 6707 is a burst with a measured redshift of 0.0085, yet a relatively low fluence of $4 \times 10^{-6}\,\rm{erg}/\rm{cm}^2$. Together, this implies a luminosity on the order of $10^{45}\,$erg/s, well below the normal range considered.  Our calculation of the Lorentz factor, which depends on the luminosity of the GRB, yields a value of about 25,000, much larger than the normally allowed range. Results in model 2 are, therefore, not particularly realistic for this particular burst (the rates are actually very low for this model). The other models are only affected by this in the calculation of the synchrotron energy loss scale.

\item[(ii)] GRBs 7648, 6891 and 1997 each have Lorentz factors below 100.  In all three cases, the burst has been found to have a low luminosity, $\sim 10^{49}\,$erg/s, which contributes to this result.  The results of models 2, 3 or 4 are only affected by this in their low synchrotron energy loss scales, and are, therefore, conservative. Results for model 1 should be interpreted with caution for these three GRBs.

\item[(iii)] GRBs 1025 and 8086, both long duration bursts, have been found to have very low variabilities and, therefore, low luminosities. This results in very high Lorentz factors in our calculation (20,000 and 5,000, respectively). Given the uncertainties involved in the variability calculation and the variability-luminosity relationship, we feel that these luminosities and Lorentz factors are unlikely to accurately represent these bursts. 

\end{itemize}

\section{Results and Discussion}

As already mentioned, our results are summarized in a series of tables. They summarize our fits to the BATSE data that provide the input spectrum for calculating the neutrino emission. The neutrino event rates for the 4 classes and the 4 models are also tabulated (see tables 1 through 8). In tables 9-12, we summarize the event rates for a generic kilometer-scale telescope, such as IceCube, for the 4 classes of GRBs. For the duration and in the direction of a GRB, the background in a neutrino telescope should be negligible. Therefore individual events represent a meaningful observation when coincident with a GRB.  We next evaluate the prospects for such observations.

The AMANDA experiment has operated for approximately four years (1997-2000) that overlap with the BATSE mission (Barouch \& Hardtke 2001). Data has been collected for several hundred GRBs with an effective telescope area of order 5,000 m$^2$ . AMANDA-II, 
the completed version of the experiment, with approximately 50,000 m$^2$ effective area for the high energy neutrinos emitted by GRBs, was commissioned less than half a year before the end of the BATSE mission. Nevertheless, of order 100 bursts occurred during that  period for which coincident observations were made. With effective areas significantly below the reference square kilometer of future neutrino telescopes, AMANDA is not expected to test the GRB models considered here. For long duration GRBs, the most common classification, we anticipate on the order of 0.01 events (muons+showers) per square kilometer per GRB for models 1 and 2. For AMANDA-II, with one twentieth of this area, and only capable of observing northern hemisphere GRBs ($\sim 500$/yr), we predict on the order of .3 events (muons+showers) per year of observation. AMANDA-B10, with smaller area, should observe one tenth of this rate. It is interesting, however, to note that such experiments are on the threshold of observation at this time. IceCube with a square kilometer of effective area, now under construction, will likely cross this threshold.

For the high energies considered here, IceCube should be able to make observations of 1000 bursts over $4\pi$ steradians during one year. Where models 1 and 2 are concerned, we expect that an event (muon or shower) will be observed from roughly 10 GRBs. For models 3 and 4 we predict $\sim 7$ events per year for models with $t_{\rm{sd}}\simeq 0.07$ yr (model 4) , but only around one event per ten years if $t_{\rm{sd}}$ is somewhat larger, such as 0.4 yr (model 3). The prospects for observation depends strongly on $t_{\rm{sd}}$, as expected.

We expect that classes of GRBs with lower average fluence, such as short duration GRBs and X-ray flash candidates, will be more difficult to observe. X-ray flash candidates, although unlikely to be observable in models 1 and 2, could possibly be observed 
in models 3 and 4. If $t_{\rm{sd}}\simeq 0.07$ yr (model 4), we predict order one event (muon or shower) for every 1000 bursts. With only 
$\sim 100$ X-ray flashes thought to occur per year, such an observation will be unlikely. If such events are more common, however, they may contribute significantly to the diffuse high energy neutrino flux. Observations of neutrinos from low peak energy GRBs would be strong evidence for a supranova progenitor model.

\begin{figure}
\includegraphics[width=14cm]{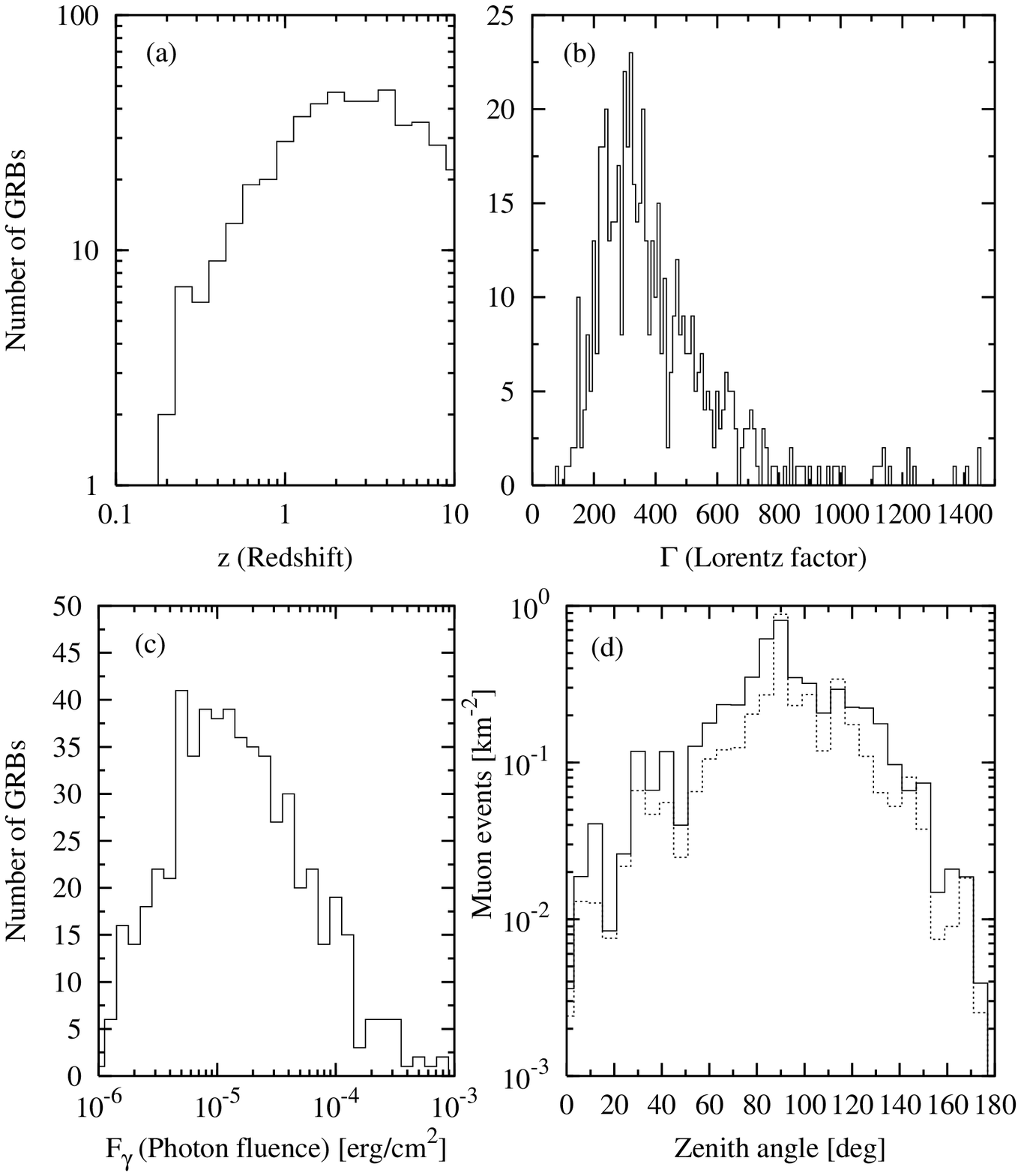}
\caption{\label{figdist}
The distribution of (a) redshifts, (b) Lorentz factors, (c) fluences and (d) zenith angles (for muon events: dotted line corresponds to model 1, solid line to model 2) for long duration GRBs. In (a), the decrease at large redshift is due to sampling bias.
 The Lorentz factors in (b) were calculated as described by Eqs.~\ref{eq:gamma} and ~\ref{eq:epeak}.  Frame (d) demonstrates the advantages of long muon range and poor absorption near the horizon for muon track detection.}
\end{figure}

\begin{figure}
\includegraphics[width=14cm]{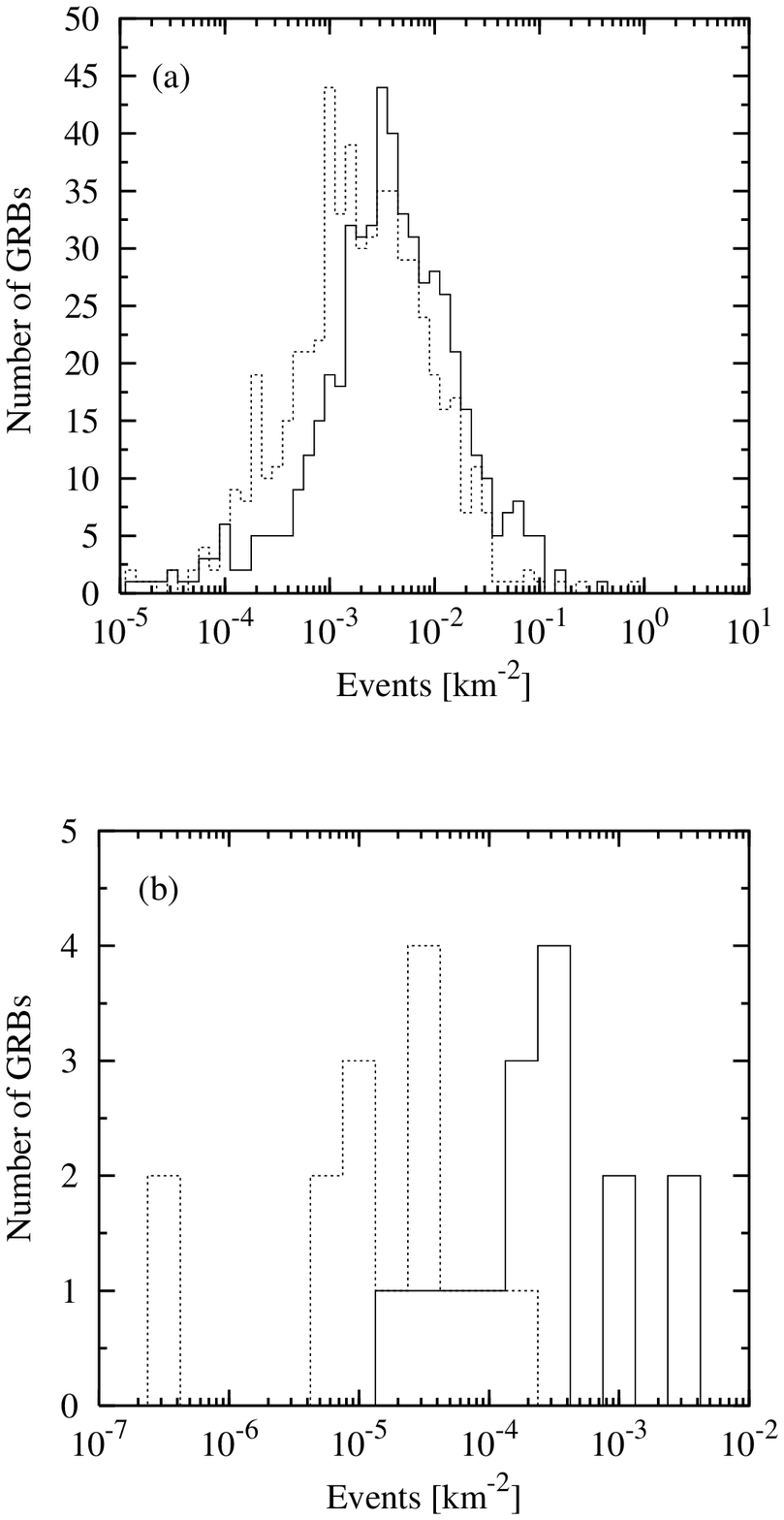}
\caption{\label{figevn}
The distribution of the estimated number of muon events from individual (a) long duration GRBs (the dotted line corresponds to model 1 and the solid line to model 2) and (b) X-ray flash candidates (dotted line corresponds to model 3 and the solid line to model 4). 
Note that the majority of events result from a relatively small number of GRBs.}
\end{figure}

It is interesting to note that the majority of the neutrino events from GRBs come from a relatively small fraction of the GRB population (Alvarez, Halzen \& 
Hooper 2000; Halzen \& Hooper 1999). 
In Figure \ref{figdist} we show some of the factors which go into this 
conclusion.  The occasional nearby (low redshift), large $f_{\pi}$, high 
fluence and/or near horizontal burst can dominate the event rate calculation.  The distribution of the number of events per burst (or X-ray flash candidate) is shown in Figure \ref{figevn}. 

In summary, taking advantage of the large body of GRB statistics available in the BATSE catalog, we have attempted to estimate the neutrino fluxes and event rates in neutrino telescopes associated with GRBs for a variety of theoretical models. Our analysis has yielded several conclusions:

\begin{itemize}

\item[(i)] Gamma-ray bursts with high fluence, most often long duration bursts, provide the best opportunity for neutrino observations.

\item[(ii)] For typical gamma-ray bursts, proton interactions with fireball photons provides the largest neutrino signal. We have also shown that the rates are relatively model independent.

\item[(iii)] For gamma-ray bursts with very low peak energies, possibly associated with X-ray flashes, very little energy is transfered into pions (and, therefore, neutrinos) in interactions with fireball photons.  Interactions with a surrounding pulsar wind bubble, however, can yield interesting neutrino fluxes. This is an illustration that observation of neutrinos is likely to help 
decipher the progenitor mechanism.

\item[(iv)] While our calculations indicate that existing neutrino telescopes, such as AMANDA, are not likely to have the sensitivity to observe gamma-ray burst neutrinos, next generation, kilometer-scale observatories, such as IceCube, will be capable of
 observing on the order of ten bursts each year.

\end{itemize}

\acknowledgements

We would like to thank Jonathan Granot, Rellen Hardtke,
Robert Preece, Ricardo V\'azquez and Eli Waxman for valuable discussions. 
This research was supported  by the U.S.~Department of Energy
under grant DE-FG02-95ER40896
and by the Wisconsin Alumni Research Foundation. 
J.A.-M. is supported by MCYT (FPA 2001-3837).

\newpage

 \begin{table}
 \caption{Characteristics for GRBs with redshifts 
 measured from optical afterglow observations.
 The fluence and zenith angle of each burst are 
 taken from the BATSE catalog, whereas the 
 luminosity is derived from the redshift and 
 fluence. The break energy is obtained by a fit to the BATSE data 
 and the Lorentz factor is calculated as described 
 in Appendix A. Note that GRB 6707 is an anomalous 
 burst (see \S 6).
 \label{table:I}
 }
 \begin{tabular}{c c c c c c c c c} \hline
 &~BATSE $\#$~&~$F_{\gamma}$(erg/$\rm{cm}^2$)
 ~&~$L_{\gamma}$(erg/s)~&~$z$~&~
 $E_{\gamma}^b$(MeV)~&~$\Gamma$~&~$\theta_z$
 (degrees)~&\\
 \hline \hline
&8079&$1.61~10^{-6}$~&~$4.9~10^{51}$~&~~1.118&0.460&~   222.&166.2&~\\
&7906&$2.51~10^{-4}$~&~$2.0~10^{52}$~&~~1.020&1.460&~   281.&100.9&~\\
&7648&$5.83~10^{-6}$~&~$2.0~10^{49}$~&~~0.434&0.710&~    89.&101.8&~\\
&7560&$2.06~10^{-5}$~&~$2.0~10^{51}$~&~~1.619&0.100&~   925.&  9.8&~\\
&7549&$2.11~10^{-4}$~&~$1.2~10^{51}$~&~~1.300&  1  &~   176.& 63.8&~\\
&7343&$4.87~10^{-4}$~&~$4.0~10^{51}$~&~~1.600&0.280&~   214.&132.3&~\\
&6891&$6.22~10^{-5}$~&~$1.5~10^{49}$~&~~0.966&0.280&~    85.&102.0&~\\
&6707&$4.01~10^{-6}$~&~$1.3~10^{45}$~&~~0.009&0.430&~ 25299.& 36.9&~\\
&6533&$1.25~10^{-5}$~&~$2.0~10^{52}$~&~~3.420&0.460&~   281.&156.2&~\\
&6225&$3.96~10^{-6}$~&~$1.2~10^{50}$~&~~0.835&1.400&~   119.&170.6&~\\
 \hline \hline
 \end{tabular}
 \end{table}
 \vspace{1.3cm}
 \begin{table}
 \caption{Estimated neutrino event rates for GRBs with 
 redshifts 
 measured from optical afterglow observations.
 Rates for muon and shower events are shown for 
 the IceCube neutrino telescope. Tau events are 
 not shown. Note that GRB 6707 is an anomalous burst.
 Model 1 and model 2 correspond to the internal
shock scenario. In model 1 we fix the fraction of energy 
transferred from protons to pions to a constant value $f_\pi=20\%$, 
and in model 2 we calculate $f_\pi$ from the observed GRB 
characteristics using Eq.(\ref{eq:fpi2}). 
Models 3 and 4 correspond to the supranova scenario, with choices
of the relevant parameter $t_{\rm sd}=0.4$ and 0.07 years 
respectively. Note that in the supranova scenario (models 3 or 4), internal shock processes (models 1 or 2) also contribute to the event rate. The total predicted rate for models 3 or 4 is, therefore, the sum of rates shown for the supranova model and the rates shown for the internal shock model.
 \label{table:II}
 }
 \begin{tabular} {c c c c c c c c c c} \hline
 & \multicolumn{2}{c }{Model 1} & 
 \multicolumn{2}{c }{Model 2} & 
 \multicolumn{2}{c }{Model 3} & 
 \multicolumn{2}{c }{Model 4} & \\
 \hline
 BATSE $\#$&$\mu$&Shower
 &$\mu$&Shower&
 $\mu$&Shower
 &$\mu$&Shower&\\
 \hline \hline
8079&$3.1~10^{-5}$&$6.8~10^{-6}$&$5.2~10^{-5}$&$1.2~10^{-5}$&$1.7~10^{-8}$&$2.9~10^{-9}$&$3.1~10^{-6}$&$5.8~10^{-7}$&\\
7906&$9.9~10^{-2}$&$2.1~10^{-2}$&$9.7~10^{-2}$&$2.1~10^{-2}$&$9.1~10^{-5}$&$1.1~10^{-5}$&$1.3~10^{-2}$&$1.8~10^{-3}$&\\
7648&$2.4~10^{-3}$&$5.7~10^{-4}$&$7.3~10^{-4}$&$1.8~10^{-4}$&$1.2~10^{-7}$&$1.5~10^{-8}$&$1.9~10^{-5}$&$2.7~10^{-6}$&\\
7560&$2.6~10^{-4}$&$3.0~10^{-4}$&$2.7~10^{-6}$&$3.1~10^{-6}$&$2.3~10^{-4}$&$2.7~10^{-4}$&$8.6~10^{-4}$&$9.9~10^{-4}$&\\
7549&$7.2~10^{-2}$&$4.1~10^{-2}$&$3.8~10^{-2}$&$2.1~10^{-2}$&$1.6~10^{-5}$&$8.1~10^{-6}$&$1.8~10^{-3}$&$9.3~10^{-4}$&\\
7343&$3.0~10^{-2}$&$5.3~10^{-3}$&$6.0~10^{-2}$&$1.1~10^{-2}$&$2.4~10^{-5}$&$3.5~10^{-6}$&$4.2~10^{-3}$&$6.8~10^{-4}$&\\
6891&$2.3~10^{-2}$&$4.3~10^{-3}$&$1.2~10^{-2}$&$2.2~10^{-3}$&$1.3~10^{-6}$&$1.8~10^{-7}$&$2.1~10^{-4}$&$3.2~10^{-5}$&\\
6707&$7.8~10^{-7}$&$7.4~10^{-7}$&$5.6~10^{-21}$&$5.3~10^{-21}$&$1.7~10^{-4}$&$1.5~10^{-4}$&$6.7~10^{-4}$&$6.7~10^{-4}$&\\
6533&$1.2~10^{-3}$&$3.7~10^{-4}$&$1.6~10^{-3}$&$5.1~10^{-4}$&$6.7~10^{-7}$&$1.0~10^{-7}$&$1.2~10^{-4}$&$2.1~10^{-5}$&\\
6225&$6.7~10^{-4}$&$2.8~10^{-4}$&$1.5~10^{-4}$&$6.2~10^{-5}$&$1.0~10^{-8}$&$1.8~10^{-9}$&$2.0~10^{-6}$&$3.9~10^{-7}$&\\
 \hline \hline
 \end{tabular}
 \end{table}

 \begin{table}
 \caption{Characteristics for a subsample of long 
 duration 
 GRBs extracted from the sample of 566 long duration GRBs. 
The luminoisty of each burst is 
 calculated from its variability, as described 
 in the text.
 The fluence and zenith angle of each burst are 
 taken from the BATSE catalog, wheras the 
 redshift is derived from the luminosity and 
 fluence. The break energy is obtained by fitting the BATSE data, 
 and the Lorentz factor is calculated as described 
 in Appendix A.
 \label{table:III}
 }
 \begin{tabular}{c c c c c c c c c} \hline
 &~BATSE $\#$~&~$F_{\gamma}$(erg/$\rm{cm}^2$)
 ~&~$L_{\gamma}$(erg/s)~&~$z$~&~
 $E_{\gamma}^b$(MeV)~&~$\Gamma$~&~$\theta_z$
 (degrees)~&\\
 \hline \hline
& 676&$4.19~10^{-5}$~&~$1.7~10^{52}$~&~~1.883&0.260&~   273.&135.2&~\\
&1606&$6.20~10^{-5}$~&~$1.3~10^{52}$~&~~1.313&0.280&~   261.& 45.2&~\\
&2102&$2.47~10^{-6}$~&~$4.6~10^{50}$~&~~0.644&0.330&~   150.& 34.7&~\\
&2431&$1.84~10^{-5}$~&~$1.1~10^{51}$~&~~0.251&0.480&~   173.& 71.0&~\\
&2586&$1.45~10^{-5}$~&~$2.9~10^{53}$~&~~4.487&0.500&~   437.& 99.3&~\\
&2798&$2.31~10^{-4}$~&~$1.9~10^{51}$~&~~0.381&0.460&~   190.& 30.0&~\\
&3356&$1.60~10^{-6}$~&~$7.9~10^{51}$~&~~1.837&0.230&~   241.& 65.8&~\\
&5644&$1.22~10^{-5}$~&~$9.5~10^{52}$~&~~3.376&0.380&~   364.&131.9&~\\
&6397&$2.11~10^{-5}$~&~$9.3~10^{51}$~&~~1.306&0.490&~   247.& 81.6&~\\
&6672&$7.81~10^{-6}$~&~$1.6~10^{53}$~&~~4.311&0.250&~   398.&108.3&~\\
&7822&$9.09~10^{-6}$~&~$2.3~10^{52}$~&~~2.417&0.420&~   287.&134.0&~\\
&8008&$1.07~10^{-4}$~&~$2.8~10^{50}$~&~~0.255&1.940&~   138.&144.3&~\\
 \hline \hline
 \end{tabular}
 \end{table}
 \vspace{1.3cm}
 \begin{table}
 \caption{Estimated neutrino event rates for the sample of long 
 duration GRBs described in table \ref{table:III}.
 Rates for muon and shower events are shown for 
 the IceCube neutrino telescope. Tau events are 
 not shown. Note that in the supranova scenario (models 3 or 4), internal
 shock processes (models 1 or 2) also contribute to the event rate. The total predicted rate for models 3 or 4 is, therefore, the sum of rates shown for the supranova model and the rates shown for the internal shock model.
 \label{table:IV}
 }
 \begin{tabular} {c c c c c c c c c c} \hline
 & \multicolumn{2}{c }{Model 1~} & 
 \multicolumn{2}{c }{Model 2~} & 
 \multicolumn{2}{c }{Model 3~} & 
 \multicolumn{2}{c }{Model 4~} & \\
 \hline
 BATSE $\#$&$\mu$&Shower
 &$\mu$&Shower&
 $\mu$&Shower
 &$\mu$&Shower&\\
 \hline \hline
 676&$3.1~10^{-3}$&$6.9~10^{-4}$&$8.5~10^{-3}$&$1.9~10^{-3}$&$3.6~10^{-6}$&$5.1~10^{-7}$&$6.1~10^{-4}$&$9.7~10^{-5}$&\\
1606&$4.8~10^{-3}$&$4.1~10^{-3}$&$1.4~10^{-2}$&$1.2~10^{-2}$&$6.6~10^{-6}$&$5.4~10^{-6}$&$8.3~10^{-4}$&$6.8~10^{-4}$&\\
2102&$2.6~10^{-4}$&$2.6~10^{-4}$&$3.8~10^{-4}$&$3.7~10^{-4}$&$5.5~10^{-8}$&$5.3~10^{-8}$&$7.3~10^{-6}$&$7.0~10^{-6}$&\\
2431&$1.6~10^{-3}$&$6.0~10^{-4}$&$2.6~10^{-3}$&$1.0~10^{-3}$&$1.3~10^{-6}$&$4.9~10^{-7}$&$1.5~10^{-4}$&$5.8~10^{-5}$&\\
2586&$2.9~10^{-3}$&$6.5~10^{-4}$&$6.3~10^{-3}$&$1.4~10^{-3}$&$3.1~10^{-5}$&$4.1~10^{-6}$&$3.1~10^{-2}$&$3.9~10^{-3}$&\\
2798&$1.9~10^{-2}$&$2.0~10^{-2}$&$3.5~10^{-2}$&$3.7~10^{-2}$&$8.1~10^{-4}$&$8.2~10^{-6}$&$9.4~10^{-4}$&$9.4~10^{-4}$&\\
3405&$1.3~10^{-3}$&$1.1~10^{-3}$&$2.5~10^{-3}$&$2.2~10^{-3}$&$1.7~10^{-5}$&$1.5~10^{-5}$&$1.3~10^{-2}$&$1.1~10^{-2}$&\\
5644&$1.4~10^{-3}$&$3.5~10^{-4}$&$3.3~10^{-3}$&$8.3~10^{-4}$&$3.1~10^{-6}$&$4.6~10^{-7}$&$4.8~10^{-4}$&$8.2~10^{-5}$&\\
6397&$4.3~10^{-3}$&$8.1~10^{-4}$&$7.6~10^{-3}$&$1.5~10^{-3}$&$9.0~10^{-6}$&$1.6~10^{-6}$&$1.1~10^{-3}$&$2.0~10^{-4}$&\\
6672&$1.9~10^{-3}$&$2.8~10^{-4}$&$6.0~10^{-3}$&$8.6~10^{-4}$&$8.4~10^{-6}$&$1.1~10^{-6}$&$9.9~10^{-4}$&$1.6~10^{-4}$&\\
7822&$9.7~10^{-4}$&$1.7~10^{-4}$&$1.8~10^{-3}$&$3.2~10^{-4}$&$1.0~10^{-6}$&$1.5~10^{-7}$&$1.8~10^{-4}$&$2.9~10^{-5}$&\\
8008&$2.4~10^{-2}$&$8.7~10^{-3}$&$7.4~10^{-3}$&$2.7~10^{-3}$&$8.4~10^{-7}$&$1.1~10^{-7}$&$1.7~10^{-4}$&$2.5~10^{-5}$&\\
 \hline \hline
 \end{tabular}
 \end{table}

 \begin{table}
 \caption{Characteristics for a subsample of short 
 duration GRBs extracted from the larger sample of 199
short duration GRBs.
 The redshift of each event is 
 fixed to $z=1$.
 The fluence and zenith angle of each burst are 
 taken from the BATSE catalog, whereas the 
 luminosity is derived from the redshift and 
 fluence. The break energy is obtained by fitting the BATSE data, 
 and the Lorentz factor is calculated as described 
 in the Appendix A.
 \label{table:V}
 }
 \begin{tabular}{c c c c c c c c c} \hline
 &~BATSE $\#$~&~$F_{\gamma}$(erg/$\rm{cm}^2$)
 ~&~$L_{\gamma}$(erg/s)~&~$z$~&~
 $E_{\gamma}^b$(MeV)~&~$\Gamma$~&~$\theta_z$
 (degrees)~&\\
 \hline \hline
& 603&$8.93~10^{-7}$~&~$1.5~10^{52}$~&~~1&0.480&~   392.& 54.2&~\\
&1073&$2.89~10^{-7}$~&~$2.4~10^{52}$~&~~1&0.490&~   425.&101.2&~\\
&1518&$1.32~10^{-6}$~&~$7.4~10^{51}$~&~~1&0.240&~  1558.& 41.3&~\\
&2312&$5.75~10^{-7}$~&~$2.6~10^{52}$~&~~1&0.280&~   431.& 72.2&~\\
&2861&$1.77~10^{-6}$~&~$4.0~10^{51}$~&~~1&0.390&~   315.&130.8&~\\
&3215&$1.38~10^{-6}$~&~$4.5~10^{52}$~&~~1&0.270&~   471.&106.6&~\\
&3940&$3.66~10^{-7}$~&~$1.2~10^{52}$~&~~1&0.080&~  2390.& 71.1&~\\
&5212&$7.35~10^{-7}$~&~$2.5~10^{52}$~&~~1&0.270&~   427.& 28.0&~\\
&6299&$2.14~10^{-7}$~&~$3.6~10^{52}$~&~~1&0.180&~  1208.& 35.7&~\\
&7063&$1.41~10^{-6}$~&~$1.0~10^{53}$~&~~1&0.280&~   541.& 36.3&~\\
&7447&$2.86~10^{-7}$~&~$2.5~10^{52}$~&~~1&0.790&~   427.& 63.1&~\\
&7980&$2.15~10^{-7}$~&~$3.2~10^{52}$~&~~1&0.070&~  2006.&  1.6&~\\
 \hline \hline
 \end{tabular}
 \end{table}
 \vspace{1.3cm}
 \begin{table}
 \caption{Estimated neutrino event rates for the sample of short 
 duration GRBs described in table \ref{table:V}.
 Rates for muon and shower events are shown for 
 the IceCube neutrino telescope. Tau events are 
 not shown. Note that in the supranova scenario (models 3 or 4), internal
 shock processes (models 1 or 2) also contribute to the event rate. The total predicted rate for models 3 or 4 is, therefore, the sum of rates shown for the supranova model and the rates shown for the internal shock model.
 \label{table:VI}
 }
 \begin{tabular} {c c c c c c c c c c} \hline
 & \multicolumn{2}{c }{Model 1~} & 
 \multicolumn{2}{c }{Model 2~} & 
 \multicolumn{2}{c }{Model 3~} & 
 \multicolumn{2}{c }{Model 4~} & \\
 \hline
 BATSE $\#$&$\mu$&Shower
 &$\mu$&Shower&
 $\mu$&Shower
 &$\mu$&Shower&\\
 \hline \hline
 603&$6.2~10^{-5}$&$4.4~10^{-5}$&$2.3~10^{-4}$&$1.6~10^{-4}$&$1.9~10^{-8}$&$1.3~10^{-8}$&$2.6~10^{-6}$&$1.8~10^{-6}$&\\
1073&$2.3~10^{-5}$&$3.8~10^{-6}$&$9.0~10^{-5}$&$1.5~10^{-5}$&$2.0~10^{-8}$&$2.6~10^{-9}$&$3.0~10^{-6}$&$4.4~10^{-7}$&\\
1518&$5.6~10^{-6}$&$4.9~10^{-6}$&$1.5~10^{-7}$&$1.3~10^{-7}$&$1.5~10^{-6}$&$1.3~10^{-6}$&$5.1~10^{-5}$&$4.5~10^{-5}$&\\
2312&$2.3~10^{-5}$&$8.4~10^{-6}$&$1.1~10^{-4}$&$3.9~10^{-5}$&$3.0~10^{-8}$&$1.1~10^{-8}$&$4.0~10^{-6}$&$1.4~10^{-6}$&\\
2861&$5.1~10^{-5}$&$8.3~10^{-6}$&$1.7~10^{-4}$&$2.7~10^{-5}$&$1.5~10^{-8}$&$2.2~10^{-9}$&$2.9~10^{-6}$&$4.8~10^{-7}$&\\
3215&$7.5~10^{-5}$&$1.3~10^{-5}$&$3.6~10^{-4}$&$6.2~10^{-5}$&$9.1~10^{-8}$&$1.2~10^{-8}$&$1.4~10^{-5}$&$2.1~10^{-6}$&\\
3940&$2.9~10^{-6}$&$1.1~10^{-6}$&$6.6~10^{-8}$&$2.6~10^{-8}$&$6.8~10^{-6}$&$2.6~10^{-6}$&$3.6~10^{-5}$&$1.4~10^{-5}$&\\
5212&$2.3~10^{-5}$&$2.4~10^{-5}$&$1.1~10^{-4}$&$1.1~10^{-4}$&$1.3~10^{-8}$&$1.3~10^{-8}$&$1.8~10^{-6}$&$1.8~10^{-6}$&\\
6299&$1.6~10^{-6}$&$1.6~10^{-6}$&$7.4~10^{-7}$&$7.0~10^{-7}$&$1.0~10^{-7}$&$9.9~10^{-8}$&$1.3~10^{-5}$&$1.2~10^{-5}$&\\
7063&$8.5~10^{-6}$&$8.0~10^{-6}$&$4.2~10^{-5}$&$4.0~10^{-5}$&$5.0~10^{-8}$&$4.7~10^{-8}$&$5.8~10^{-6}$&$5.5~10^{-6}$&\\
7447&$3.7~10^{-5}$&$2.2~10^{-5}$&$1.1~10^{-4}$&$6.6~10^{-5}$&$9.9~10^{-9}$&$5.2~10^{-9}$&$1.3~10^{-6}$&$7.0~10^{-7}$&\\
7980&$3.6~10^{-6}$&$4.4~10^{-6}$&$5.0~10^{-7}$&$6.0~10^{-7}$&$3.1~10^{-7}$&$3.7~10^{-7}$&$6.2~10^{-6}$&$7.2~10^{-6}$&\\
 \hline \hline
 \end{tabular}
 \end{table}

 \begin{table}
 \caption{Characteristics for X-ray flash 
 candidates. The redshift to each event is 
 fixed to $z=1$.
 The fluence and zenith angle of each burst are 
 taken from the BATSE catalog, whereas the 
 luminosity is derived from the redshift and 
 fluence. The break energy is obtained by fitting the BATSE data, 
 and the Lorentz factor is calculated as described 
 in Appendix A.
 \label{table:VII}
 }
 \begin{tabular}{c c c c c c c c c} \hline
 &~BATSE $\#$~&~$F_{\gamma}$(erg/$\rm{cm}^2$)
 ~&~$L_{\gamma}$(erg/s)~&~$z$~&~
 $E_{\gamma}^b$(MeV)~&~$\Gamma$~&~$\theta_z$
 (degrees)~&\\
 \hline \hline
& 659&$7.60~10^{-6}$~&~$9.9~10^{48}$~&~1~&0.140&~  1508.&109.4&~\\
& 717&$5.44~10^{-7}$~&~$1.0~10^{49}$~&~1~&0.050&~  2505.& 98.1&~\\
& 927&$6.31~10^{-7}$~&~$1.7~10^{49}$~&~1~&0.130&~  1357.&128.4&~\\
&1244&$3.12~10^{-6}$~&~$1.9~10^{49}$~&~1~&0.060&~  1961.&133.7&~\\
&2381&$4.96~10^{-7}$~&~$1.1~10^{50}$~&~1~&0.050&~  1367.& 79.8&~\\
&2917&$5.27~10^{-7}$~&~$3.8~10^{49}$~&~1~&0.160&~  1004.& 94.6&~\\
&2990&$4.78~10^{-7}$~&~$5.3~10^{49}$~&~1~&0.050&~  1654.&101.6&~\\
&3141&$1.55~10^{-7}$~&~$5.4~10^{50}$~&~1~&0.050&~   926.&141.7&~\\
&5483&$4.18~10^{-7}$~&~$6.4~10^{50}$~&~1~&0.090&~   663.&154.8&~\\
&5497&$3.21~10^{-6}$~&~$7.1~10^{48}$~&~1~&0.060&~  2500.&141.1&~\\
&5627&$3.16~10^{-7}$~&~$8.4~10^{49}$~&~1~&0.070&~  1247.& 63.4&~\\
&6167&$9.21~10^{-6}$~&~$2.3~10^{49}$~&~1~&0.070&~  1724.&126.3&~\\
&6437&$4.56~10^{-7}$~&~$1.2~10^{50}$~&~1~&0.050&~  1359.&118.4&~\\
&6582&$1.17~10^{-6}$~&~$4.8~10^{50}$~&~1~&0.060&~   872.& 27.7&~\\
&7942&$9.86~10^{-7}$~&~$4.5~10^{49}$~&~1~&0.050&~  1722.&134.5&~\\
 \hline \hline
 \end{tabular}
 \end{table}
 \vspace{1.3cm}
 \begin{table}
 \caption{Estimated neutrino event rates for X-ray flash 
 candidates.
 Rates for muon and shower events are shown for 
 the IceCube neutrino telescope. Tau events are 
 not shown. Note that in the supranova scenario (models 3 or 4), internal 
shock processes (models 1 or 2) also contribute to the event rate. The total predicted rate for models 3 or 4 is, therefore, the sum of rates shown for the supranova model and the rates shown for the internal shock model.
 \label{table:VIII}
 }
 \begin{tabular} {c c c c c c c c c c} \hline
 & \multicolumn{2}{c }{Model 1~} & 
 \multicolumn{2}{c }{Model 2~} & 
 \multicolumn{2}{c }{Model 3~} & 
 \multicolumn{2}{c }{Model 4~} & \\
 \hline
 BATSE $\#$&$\mu$&Shower
 &$\mu$&Shower&
 $\mu$&Shower
 &$\mu$&Shower&\\
 \hline \hline
 659&$4.8~10^{-5}$&$4.6~10^{-6}$&$6.6~10^{-11}$&$6.3~10^{-12}$&$1.5~10^{-4}$&$1.7~10^{-5}$&$2.3~10^{-3}$&$5.1~10^{-4}$&\\
 717&$1.3~10^{-6}$&$1.7~10^{-8}$&$6.6~10^{-12}$&$8.6~10^{-15}$&$4.1~10^{-5}$&$4.8~10^{-6}$&$2.3~10^{-4}$&$5.7~10^{-5}$&\\
 927&$2.3~10^{-5}$&$6.8~10^{-6}$&$9.1~10^{-11}$&$2.7~10^{-11}$&$3.6~10^{-6}$&$3.9~10^{-7}$&$1.2~10^{-4}$&$2.8~10^{-5}$&\\
1244&$9.1~10^{-7}$&$7.0~10^{-8}$&$1.9~10^{-13}$&$1.5~10^{-13}$&$2.5~10^{-5}$&$3.1~10^{-6}$&$6.6~10^{-4}$&$1.8~10^{-4}$&\\
2381&$1.1~10^{-5}$&$2.4~10^{-6}$&$7.3~10^{-10}$&$1.6~10^{-10}$&$2.4~10^{-5}$&$5.1~10^{-6}$&$1.8~10^{-4}$&$4.4~10^{-5}$&\\
2917&$8.6~10^{-5}$&$1.8~10^{-5}$&$2.0~10^{-9}$&$4.3~10^{-10}$&$2.4~10^{-5}$&$2.3~10^{-6}$&$1.9~10^{-4}$&$3.61~0^{-5}$&\\
2990&$1.4~10^{-6}$&$1.1~10^{-7}$&$2.0~10^{-11}$&$1.5~10^{-12}$&$1.8~10^{-5}$&$2.0~10^{-6}$&$1.7~10^{-4}$&$3.8~10^{-5}$&\\
3141&$4.0~10^{-7}$&$7.3~10^{-8}$&$5.9~10^{-10}$&$1.1~10^{-10}$&$2.6~10^{-7}$&$2.7~10^{-8}$&$1.7~10^{-5}$&$3.7~10^{-6}$&\\
5483&$6.7~10^{-7}$&$7.4~10^{-8}$&$2.5~10^{-9}$&$2.7~10^{-10}$&$2.5~10^{-7}$&$2.3~10^{-8}$&$1.8~10^{-5}$&$3.9~10^{-6}$&\\
5497&$4.5~10^{-7}$&$3.5~10^{-8}$&$1.4~10^{-13}$&$1.0~10^{-14}$&$2.8~10^{-5}$&$3.9~10^{-6}$&$6.8~10^{-4}$&$2.1~10^{-4}$&\\
5627&$4.9~10^{-6}$&$2.6~10^{-6}$&$2.4~10^{-10}$&$1.3~10^{-10}$&$5.8~10^{-6}$&$3.0~10^{-6}$&$5.1~10^{-5}$&$2.7~10^{-5}$&\\
6167&$5.4~10^{-6}$&$4.1~10^{-7}$&$2.0~10^{-11}$&$1.5~10^{-12}$&$8.1~10^{-5}$&$9.6~10^{-6}$&$2.0~10^{-3}$&$5.2~10^{-4}$&\\
6437&$4.5~10^{-7}$&$3.5~10^{-8}$&$3.1~10^{-11}$&$2.4~10^{-12}$&$4.2~10^{-6}$&$4.6~10^{-7}$&$1.0~10^{-4}$&$2.4~10^{-5}$&\\
6582&$1.2~10^{-5}$&$1.2~10^{-5}$&$1.6~10^{-8}$&$1.7~10^{-8}$&$8.5~10^{-6}$&$8.9~10^{-6}$&$8.0~10^{-5}$&$8.2~10^{-5}$&\\
7942&$2.6~10^{-7}$&$1.7~10^{-8}$&$2.7~10^{-12}$&$1.7~10^{-13}$&$6.3~10^{-6}$&$7.5~10^{-7}$&$1.9~10^{-4}$&$5.0~10^{-5}$&\\
 \hline \hline
 \end{tabular}
 \end{table}

 \begin{table}
 \caption{Sum of event rates and event rate per GRB 
in a kilometer scale neutrino 
 telescope for GRBs with known redshift. Note that in the supranova scenario (models 3 or 4), internal shock processes (models 1 or 2) also contribute to the event rate. The total predicted rate for models 3 or 4 is, therefore, the sum of rates shown for the supranova model and the rates shown for the internal shock model. 
 \label{table:IX}
 }
 \begin{tabular}{c c c c c c c c c} \hline
 &~Model~&~total $\mu$~&~total showers
 ~&~total $\tau$~&~$\mu$/GRB~&~
 showers/GRB~&~$\tau$/GRB~&\\
 \hline \hline
&1&$0.236$&$7.49~10^{-2}$~&~$2.05~10^{-3}$~&~~$1.57~10^{-2}$&$4.99~10^{-3}$&~$1.37~10^{-4}$&~\\
&2&$0.215$&$5.72~10^{-2}$~&~$1.26~10^{-3}$~&~~$1.44~10^{-2}$&$3.81~10^{-3}$&~$8.43~10^{-5}$&~\\
&3&$5.30~10^{-4}$&$4.45~10^{-4}$~&~$2.50~10^{-4}$~&~~$3.54~10^{-5}$&$2.97~10^{-5}$&~$1.67~10^{-5}$&~\\
&4&$2.12~10^{-2}$&$5.19~10^{-3}$~&~$4.57~10^{-4}$~&~~$1.41~10^{-3}$&$3.46~10^{-4}$&~$3.05~10^{-5}$&~\\
 \hline \hline
 \end{tabular}
 \end{table}

 \begin{table}
 \caption{Sum of event rates and event rate per GRB
in a kilometer scale neutrino telescope for long duration GRBs. Note that in the supranova scenario (models 3 or 4), internal shock processes (models 1 or 2) also contribute to the event rate. The total predicted rate for models 3 or 4 is, therefore, the sum of rates shown for the supranova model and the rates shown for the internal shock model.
 \label{table:X}
 }
 \begin{tabular}{c c c c c c c c c} \hline
 &~Model~&~total $\mu$~&~total showers
 ~&~total $\tau$~&~$\mu$/GRB~&~
 showers/GRB~&~$\tau$/GRB~&\\
 \hline \hline
&1&$3.54$&$1.09$~&~$5.18~10^{-2}$~&~~$6.16~10^{-3}$&$1.90~10^{-3}$&~$9.02~10^{-5}$&~\\
&2&$5.07$&$1.65$~&~$4.76~10^{-2}$~&~~$8.82~10^{-3}$&$2.87~10^{-3}$&~$8.29~10^{-5}$&~\\
&3&$4.26~10^{-2}$&$1.22~10^{-2}$~&~$1.41~10^{-2}$~&~~$7.41~10^{-5}$&$2.12~10^{-5}$&~$2.45~10^{-5}$&~\\
&4&$3.18$&$0.87$~&~$1.27~10^{-1}$~&~~$5.53~10^{-3}$&$1.51~10^{-3}$&~$2.20~10^{-4}$&~\\
 \hline \hline
 \end{tabular}
 \end{table}

 \begin{table}
 \caption{Sum of estimated event rates and event rate per GRB 
in a kilometer scale neutrino 
 telescope for short duration GRBs. Note that in the supranova scenario (models 3 or 4), internal shock processes (models 1 or 2) also contribute to the event rate. The total predicted rate for models 3 or 4 is, therefore, the sum of rates shown for the supranova model and the rates shown for the internal shock model. 
 \label{table:XI}
 }
 \begin{tabular}{c c c c c c c c c} \hline
 &~Model~&~total $\mu$~&~total showers
 ~&~total $\tau$~&~$\mu$/GRB~&~
 showers/GRB~&~$\tau$/GRB~&\\
 \hline \hline
&1&$2.17~10^{-2}$&$6.73~10^{-3}$~&~$2.41~10^{-4}$~&~~$1.09~10^{-4}$&$3.38~10^{-5}$&~$1.21~10^{-6}$&~\\
&2&$8.39~10^{-2}$&$2.58~10^{-2}$~&~$8.85~10^{-4}$~&~~$4.22~10^{-4}$&$1.30~10^{-4}$&~$4.45~10^{-6}$&~\\
&3&$1.18~10^{-4}$&$4.81~10^{-5}$~&~$4.42~10^{-5}$~&~~$5.94~10^{-7}$&$2.42~10^{-7}$&~$2.22~10^{-7}$&~\\
&4&$7.69~10^{-3}$&$2.04~10^{-3}$~&~$6.60~10^{-4}$~&~~$3.86~10^{-5}$&$1.02~10^{-5}$&~$3.32~10^{-6}$&~\\
 \hline \hline
 \end{tabular}
 \end{table}

 \begin{table}
 \caption{Sum of estimated event rates and event rate per GRB 
in a kilometer scale neutrino 
 telescopes for X-ray flash candidates. Note that in the supranova scenario (models 3 or 4), internal shock processes (models 1 or 2) also contribute to the event rate. The total predicted rate for models 3 or 4 is, therefore, the sum of rates shown for the supranova model and the rates shown for the internal shock model. 
 \label{table:XII}
 }
 \begin{tabular}{c c c c c c c c c} \hline
 &~Model~&~total $\mu$~&~total showers
 ~&~total $\tau$~&~$\mu$/GRB~&~
 showers/GRB~&~$\tau$/GRB~&\\
 \hline \hline
&1&$1.96~10^{-4}$&$4.74~10^{-5}$~&~$3.24~10^{-5}$~&~~$1.30~10^{-5}$&$3.16~10^{-6}$&~$2.16~10^{-6}$&~\\
&2&$2.22~10^{-8}$&$1.79~10^{-8}$~&~$1.59~10^{-8}$~&~~$1.48~10^{-9}$&$1.19~10^{-9}$&~$1.06~10^{-9}$&~\\
&3&$4.23~10^{-4}$&$6.09~10^{-5}$~&~$2.51~10^{-5}$~&~~$2.82~10^{-5}$&$4.06~10^{-6}$&~$1.67~10^{-6}$&~\\
&4&$7.00~10^{-3}$&$1.80~10^{-3}$~&~$2.14~10^{-5}$~&~~$4.67~10^{-4}$&$1.20~10^{-4}$&~$1.43~10^{-6}$&~\\
 \hline \hline
 \end{tabular}
 \end{table}

\newpage

\appendix
\section{Appendix: Photomeson Interactions of Protons With GRB Photons}\label{GRBphotons}

In this appendix, we describe the production of neutrinos in interactions of protons and photons in the GRB fireball. Protons predominantly produce the parent pions via the processes
\begin{eqnarray}
p\gamma \rightarrow \Delta \rightarrow n \pi^{+}
\end{eqnarray}
and
\begin{eqnarray}
p\gamma \rightarrow \Delta \rightarrow p \pi^{0}
\end{eqnarray}
which have very large cross sections of $\sigma_{\Delta}\sim 5\times 10^{-28}
\rm{cm}^2$.  The charged $\pi$'s subsequently decay producing charged
leptons and
neutrinos, while neutral $\pi$'s decay into high-energy photons. For the center-of-mass energy of a proton-photon interaction to exceed the threshold energy for producing the $\Delta$-resonance, the comoving
proton energy must meet the condition:
\begin{eqnarray}
\varepsilon^\prime_p \geq \frac{m_\Delta^2-m_p^2}{4 \varepsilon^\prime_\gamma}.
\end{eqnarray}
Throughout this paper, primed quantities are measured in the 
comoving frame and unprimed quantities in the observer frame. 
In the observer's frame,
\begin{equation}
\label{Eq:epb}
\varepsilon_p \geq 1.4 \times 10^{16} \frac{\Gamma^2_{2.5}}
{\varepsilon_{\gamma,{\rm MeV}}}\rm{eV},
\end{equation}
resulting in a neutrino energy
\begin{equation}
\varepsilon_{\nu}= \frac{1}{4} \langle x_{p \rightarrow \pi} \
\rangle \varepsilon_p \geq 7 \times
10^{14}\frac{\Gamma^2_{2.5}}
{\varepsilon_{\gamma,{\rm MeV}}}\rm{eV},
\end{equation}
where $\Gamma_{2.5}=\Gamma/10^{2.5}$ is the plasma expansion (bulk) 
Lorentz factor and $\varepsilon_{\gamma, \rm{MeV}}=
\varepsilon_{\gamma}/\rm{1 MeV}$ is the 
photon energy. $\langle x_{p \rightarrow \pi} \rangle \simeq 0.2$ 
is the average fraction of energy transferred from the initial proton to the
produced pion.  The factor of 1/4 is based on the estimate that the 4
final state leptons in the decay chain
$\pi^{+} \rightarrow \nu_{\mu} \mu^+ \rightarrow\nu_{\mu} e^+
\nu_e \bar{\nu_{\mu}}$ equally share the pion energy. These approximations are adequate given the uncertainties in the astrophysics of the problem.

For each proton energy, the resulting neutrino spectrum traces the
broken power law spectrum of photons which we fit to the BATSE data using the broken power law parameterization

\begin{equation}
F_{\gamma}=\varepsilon_{\gamma}dn_{\gamma}/d\varepsilon_{\gamma}
\propto\left\{ \begin{array}{ll} 
\varepsilon_{\gamma}^{-\alpha} & 
\varepsilon_{\gamma}<\varepsilon^b_{\gamma} \\ 
\varepsilon_{\gamma}^{-\beta} & \varepsilon_{\gamma}>\varepsilon^b_{\gamma}
\end{array}
\right. \;.
\label{eq:Fnu}
\end{equation}
Summing over proton energies results in a neutrino spectrum with
the same spectra slopes, $\alpha$ and $\beta$, as for the gamma-ray spectra
in the BATSE data, but with a break energy of order 1\,PeV in the observer frame:
\begin{equation}
\label{eq:enub}
\varepsilon^b_\nu = 7 \times
10^{14}\frac{1}{(1+z)^{2}}\frac{\Gamma^2_{2.5}}
{\varepsilon^b_{\gamma,{\rm MeV}}}\rm{eV}.
\end{equation}
We here explicitly introduce the dependence on source redshift, z. The highest energy pions may lose some energy via synchrotron  emission before decaying, thus reducing the energy of the decay neutrinos. The effect becomes important when the pion lifetime $\tau'_\pi\approx 2.6\times 10^{-8}\; \varepsilon'_\pi/(m_\pi c^2)\;\,$s becomes comparable to the synchrotron loss time
\begin{equation}\label{t_sync}
t'_{\rm syn}={3m_\pi^4c^3\over 4\sigma_T m_e^2 \varepsilon_\pi U'_B},
\end{equation}
where $U'_B={B'}^2/8\pi$ is the energy density of the magnetic field in the shocked fluid. $\epsilon_B$, the fraction of the internal energy carried by the magnetic field, is defined by the relation $4\pi R^2 c \Gamma^2 B'^2/8\pi=\epsilon_B L_{\rm int}$, 
where 
$R \sim 2 \Gamma^2 c t_v,$ is the collision radius. The collision radius $R$ is obtained from the consideration that different shells in the shocked fireball have velocities differing by $\Delta v\sim c/2\Gamma^2$, where $\Gamma$ is an average value representative of the entire fireball. Different shells emitted at times differing by $t_v$ therefore collide with each other after a time $t_c\sim c t_v/\Delta v$, i.e. at a radius $R = c t_c \simeq 2 \Gamma^2 c t_v$. A detailed account of the kinematics can 
be found in Halzen \& Hooper 2002. We can now compare the synchrotron loss time with the time over which the pions decay:
\begin{equation}\label{t_syn}
{t'_{\rm syn}/\tau'_\pi}= 0.21\,\epsilon_e\epsilon_B^{-1}
L_{52}^{-1}\Gamma_{2.5}^8 t_{v,-2}^2\varepsilon_{\pi18}^{-2}.
\end{equation}
Here $\epsilon_e$ is the fraction of internal energy converted to 
electrons, 
$t_{v,-2}=t_v/10^{-2}\,\,$s is the time scale of fluctuations in
the GRB lightcurve, $L_{\gamma,52}=L_{\gamma}/10^{52}\,\,$erg/s 
is the $\gamma$-ray luminosity of the GRB and
$\varepsilon_{\pi18}=\varepsilon_{\pi}/10^{18}\;$eV, is the pion energy. 
In deriving Eq.(\ref{t_syn}) we have assumed that the wind luminosity
carried by internal plasma energy, $L_{\rm int}$, is related to the 
observed $\gamma$-ray luminosity through 
$L_{\rm int}=L_{\gamma}/\epsilon_e$. This assumption is
justified because the electron synchrotron cooling time is short 
compared to the wind expansion time and hence electrons lose
all their energy radiatively.

The radiative losses become important for $t'_{\rm sync}<\tau'_\pi$,
which corresponds to $\varepsilon_{\pi}>\varepsilon_{\pi s}
\approx 4\varepsilon_{\nu s}$,
where
\begin{equation}
\label{eq:synclos}
\varepsilon^s_{\nu_{\mu}}=\frac{10^{17}}{1+z}\,
\epsilon_e^{1/2}\epsilon_B^{-1/2}L_{\gamma,52}^{-1/2}\Gamma_{2.5}^4 t_{v,-2}
\, \rm{eV}.
\end{equation}
Neutrinos from muon decay have a lifetime 100 times longer than pions,
the energy cutoff will therefore be 10 times smaller:
\begin{equation}
\label{eq:synclos2}
\varepsilon^s_{\bar{\nu}_{\mu},\nu_e}=\frac{ 10^{16}}{1+z}\,
\epsilon_e^{1/2}\epsilon_B^{-1/2}L_{\gamma,52}^{-1/2}\Gamma_{2.5}^4 t_{v,-2}
\, \rm{eV}.
\end{equation}
Above this energy, the slope of the neutrino spectrum steepens
by two to ($\beta +2$).

To normalize the neutrino spectrum to the observed GRB luminosity, we must calculate the
fraction, $f_{\pi}$, of fireball proton energy lost to pion production. The fraction of energy converted to pions is estimated from the ratio of the size of the shock, $\Delta R^\prime$, and the mean free path of a proton for photomeson interactions: 
\begin{eqnarray}
f_{\pi} \simeq \frac{\Delta R^\prime}{\lambda_{p \gamma}}\langle
x_{p\rightarrow \pi}\rangle.
\end{eqnarray}
Here, the proton mean free path is given by $\lambda_{p \gamma}=1/n_\gamma \sigma_{\Delta},$
where $n_\gamma$ is the number density of photons.
The photon number density is given by the
ratio of the photon energy density and the photon energy in the comoving
frame:
\begin{equation}
n_\gamma=\frac{U_\gamma^\prime}{\varepsilon_{\gamma}^\prime}\simeq
\bigg(\frac{L_\gamma
t_v/\Gamma}{4\pi R^2 \Delta
R^\prime}\bigg)
\bigg/
\bigg(\frac{\varepsilon_\gamma}{\Gamma}\bigg).
\end{equation}
Using these equations, and recalling that $R\simeq2 \Gamma^2 c t_v$, 
we obtain that
\begin{eqnarray}
n_\gamma\simeq\bigg(\frac{L_\gamma}{16\pi c^2 t_v
\Gamma^5 \Delta R^\prime  }\bigg)   \bigg/
\bigg(\frac{\varepsilon_\gamma}{\Gamma}\bigg)=\frac{L_\gamma} {16\pi c^2 t_v\Gamma^4 \Delta R^\prime \varepsilon_{\gamma}},
\end{eqnarray}
and the fraction of proton energy converted to $\pi$'s is 
\begin{equation}
\label{eq:fpi1}
f_{\pi} \simeq
\frac{L_{\gamma}}{\varepsilon_{\gamma}}\frac{1}{\Gamma^4 t_v}
\frac{\sigma_\Delta \langle x_{p \rightarrow \pi} \rangle}{16 \pi c^2}
\sim 0.2 \times {L_{\gamma,52}\over
\Gamma_{2.5}^4 t_{v,-2}\varepsilon_{\gamma,\rm MeV}^b}.
\end{equation}
This derivation was performed for protons at the break energy. In general,
\begin{equation}
\label{eq:fpi2}
f_\pi(\varepsilon_p)\sim0.2{L_{\gamma,52}\over
\Gamma_{2.5}^4 t_{v,-2}\varepsilon_{\gamma,\rm MeV}^b}
\times \left\{ \begin{array}{ll}
(\varepsilon_{p}/\varepsilon^b_{p})^{\alpha} &
\varepsilon_{p}>\varepsilon^b_{p} \\
(\varepsilon_{p}/\varepsilon^b_{p})^{\beta} & \varepsilon_{p}<\varepsilon^b_{p}
\end{array}
\right. \;,
\label{eq:fpi}
\end{equation}
where $\varepsilon^b_p$ is given by Eq.(\ref{Eq:epb}).

As we can see from Eq.(\ref{eq:fpi1}), $f_{\pi}$ strongly
depends on the bulk Lorentz factor $\Gamma$. It has been  pointed out by 
Halzen \& Hooper (1999) and Alvarez, Halzen \& Hooper (2000) that,
if the Lorentz factor $\Gamma$ varies significantly between bursts, then the resulting neutrino flux will be dominated 
by a few bright bursts with $f_{\pi}$ close to unity. 
However, Guetta, Spada \& Waxman (2001a) have shown that 
burst-to-burst variations in the fraction of 
fireball energy converted to neutrinos are constrained. 
First, the observational constraints imposed by $\gamma$-ray 
observations, in particular the requirement 
$\varepsilon_{\gamma}^b\geq 0.1$ MeV, imply that wind model parameters 
$(\Gamma,L,t_v$) are correlated (Guetta, Spada \& Waxman 2001b)
and that $\Gamma$ is restricted to values 
in a range much narrower than $\Delta\Gamma/\Gamma\sim1$. 
For instance, for values of $\gamma$ much smaller than average the 
fireball becomes very dense with abundant neutrino production. 
Such fireballs will also produce a thermal photon spectrum which is not 
the case for the events considered here. Second, for wind parameters that yield $f_{\pi}$ values significantly exceeding 20\%, only a small fraction of pion energy is converted to neutrinos because of pion and muon synchrotron losses as can be seen from E
q.(\ref{eq:synclos}).

We will use two methods to determine the value of the bulk Lorentz factor, $\Gamma$. For bursts with high break energies, $\varepsilon^b_{\gamma}\gtrsim 500 \, $ keV, $\Gamma$ cannot differ significantly from the minimum value for which the fireball pair 
production optical depth is $\sim 1$ near the maximum energy of $\gamma$-rays produced, $\varepsilon_{\gamma, max}$. EGRET has observed $\gamma$-rays with energies in excess of 1 GeV for six bursts, although the maximum $\gamma$-ray energy should be lower
 for the majority of GRBs. We choose 100 MeV as the default value, therefore,
\begin{equation}
\label{eq:gamma}
\Gamma\sim 250\left[L_{\gamma,52}t^{-1}_{v,-2}
\left(\frac{\varepsilon_{\gamma,max}}{100{\rm MeV}}\right)\right]^{1/6}.
\label{eq:Gammamin}
\end{equation}
Note that in the end, the value of the Lorentz factor depends weakly on 
luminosity, time structure and maximum $\gamma$-ray energy.

For bursts with lower break energies Eq.(\ref{eq:gamma}) may not be reliable because the Lorentz factor of these GRBs may be larger than estimated. 
Guetta Spada \& Waxman (2001b) have argued that the X-ray flashes identified by BeppoSAX could be produced by relativistic winds where the Lorentz factor is larger than the minimum value given in Eq.(\ref{eq:gamma}) required to produce a GRB with the characteristic photon spectrum. For GRBs with low break energy we, instead, relate the Lorentz factor to the peak energy of the $\gamma$-ray spectrum. The characteristic frequency of synchrotron emission is determined by the minimum electron Lorentz factor $\gamma_m\approx \epsilon_e (m_p/m_e)$ and by the strength of the magnetic field given above, before  Eq.(\ref{t_syn}). 
The characteristic energy of synchrotron photons,
$\varepsilon_{\gamma}^b=\Gamma h\gamma_m^2 e B'/2\pi m_e c$,
at the source redshift is
\begin{equation}
\label{eq:epeak}
\varepsilon_{\gamma}^b\approx \epsilon^{1/2}_B \epsilon_e^{3/2}
\frac{ L^{-1/2}_{\gamma,52}}{\Gamma_{2.5}^2t_{v,-2}}
{\rm MeV}.
\end{equation}
For bursts with $\varepsilon^b_{\gamma} < 500 \, $ keV we will evaluate $\Gamma$ from the break photon energy given above. At present no theory allows the determination of the values 
of the equipartition fractions $\epsilon_e$ and $\epsilon_B$.
Eq.(\ref{eq:epeak}) implies that fractions not far below unity are required
to account for the observed $\gamma$ ray emission and this is confirmed
also by simulations (Guetta Spada \& Waxman 2001b). For bursts with very large values of $\Gamma$, the peak energy is shifted 
to values lower than $40$ keV. This could be the X-ray bursts 
detected by BeppoSAX (Guetta Spada \& Waxman 2001b). From Eq.(\ref{eq:fpi1}), we estimate that the neutrino flux from such 
events is expected to be small.

\begin{figure}
\includegraphics[width=14cm]{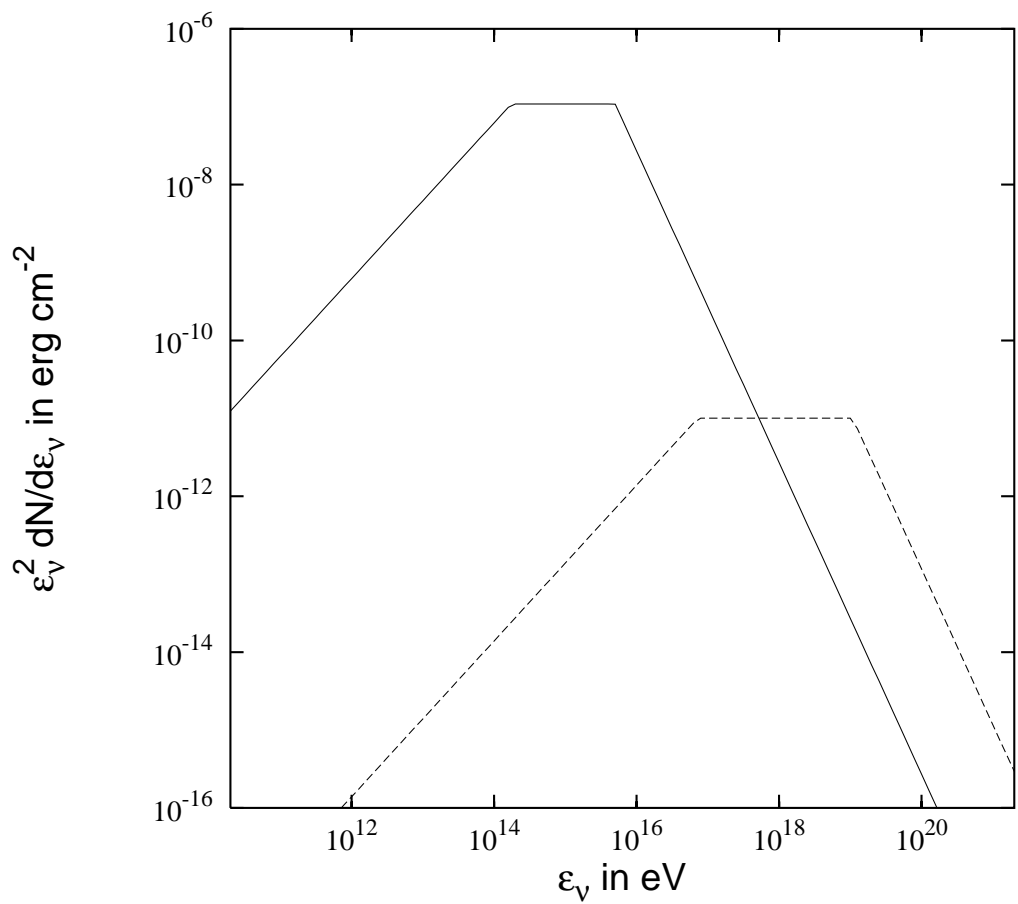}
\caption{\label{fig2a}
The muon neutrino spectrum, $\varepsilon_\nu^2(dN_\nu/d\varepsilon_\nu)$, 
for our fiducial parameters in models 1 and 2 (interactions with GRB photons).
The solid line is for a typical GRB with $\Gamma=300$ and $t_v=10$ ms, 
while the dashed line is for a 
X-ray flash candidate with $\Gamma=1440$ (calculated from Eq.~\ref{eq:epeak}) 
and $t_v=50$ ms.}
\end{figure}

We have now collected all the information to derive the neutrino from the observed $\gamma$-ray fluency $F_{\gamma}$:
\begin{eqnarray}
\frac{dN_{\nu}}{d\varepsilon_{\nu}} \varepsilon^2_{\nu} 
\simeq \frac{1}{8} \frac{1}{\epsilon_e} \frac{F_{\gamma}}
{\ln(10)} f_{\pi}.
\end{eqnarray}
BATSE detectors measure the GRB fluence $F_{\gamma}$ over two decades of photon energies, $\sim 0.02$ MeV to $\sim$ 2 MeV, corresponding to a decade of energy of the radiating electrons. The factor 1/8 takes into account that charged and neutral
pions are produced with roughly equal probabilities, and each
neutrino carries $\sim 1/4$ of the pion energy. Using Eq.(\ref{eq:fpi2}),
\begin{equation}
\varepsilon_{\nu}^2dN_{\nu}/d\varepsilon_{\nu}
\approx \frac{1}{8}\frac{1}{\epsilon_e}\frac{F_{\gamma}}{{\rm ln}(10)}
\, 0.2{L_{\gamma,52}\over
\Gamma_{2.5}^4 t_{v,-2}\varepsilon_{\gamma,\rm MeV}^b}
\times \left\{ \begin{array}{lll} 
(\varepsilon_{\nu}/\varepsilon^b_{\nu})^{\beta} & 
\varepsilon_{\nu}<\varepsilon^b_{\nu} \\ 
(\varepsilon_{\nu}/\varepsilon^b_{\nu})^{\alpha} & 
\varepsilon^b_{\nu}<\varepsilon_{\nu}<\varepsilon^s_{\nu}\\
(\varepsilon_{\nu}/\varepsilon^b_{\nu})^{\alpha} 
(\varepsilon_{\nu}/\varepsilon^s_{\nu})^{-2}& 
\varepsilon_{\nu}>\varepsilon^s_{\nu}
\end{array}
\right. \;,
\label{eq:nuflux1}
\end{equation}
where $\varepsilon^b_{\nu}$ and $\varepsilon^s_{\nu}$ are given by 
Eq.(\ref{eq:enub}), Eq.(\ref{eq:synclos}) and Eq.(\ref{eq:synclos2}). This spectrum is shown in Fig.(\ref{fig2a}).

This result depends on a number of somewhat tenuous assumptions. Simulations (GSW) actually suggest that one can simply fix $f_{\pi}\sim 0.2$ at the break energy and derive the $\nu$ flux directly from the $\gamma$-flux. Doing better may require a better understanding of the fireball phenomenology than we have now. First, the variability time may be shorter than what is observed; in most cases variability is only measured to the smallest time scale that can be detected with adequate statistics. Second, the
 parameters $\epsilon_{e}$ and $\epsilon_{B}$ are uncertain. Third, the luminosity-variability relation used to derive the luminosity for bursts with no measured redshift (see \S  \ref{zknown}) is uncertain, and in addition may have large fluctuations around the prediction. We have therefore decided to do the detailed analysis described above as well as an alternative analysis that assumes $f_{\pi}=0.2$, at the break energy, for all bursts and determines the neutrino flux directly from 
the observed gamma-ray fluence. For this alternative approach,
\begin{equation}
\frac{dN_{\nu}}{d\varepsilon_{\nu}} \varepsilon^2_{\nu} 
\simeq \frac{0.2}{8 \epsilon_e}
\frac{F_{\gamma}}{\ln(10)}\,\,
   \times  
 \left\{ \begin{array}{lll} 
(\varepsilon_{\nu}/\varepsilon^b_{\nu})^{\beta} & 
\varepsilon_{\nu}<\varepsilon^b_{\nu} \\ 
(\varepsilon_{\nu}/\varepsilon^b_{\nu})^{\alpha} & 
\varepsilon^b_{\nu}<\varepsilon_{\nu}<\varepsilon^s_{\nu}\\
(\varepsilon_{\nu}/\varepsilon^b_{\nu})^{\alpha} 
(\varepsilon_{\nu}/\varepsilon^s_{\nu})^{-2}& 
\varepsilon_{\nu}>\varepsilon^s_{\nu}
\end{array}
\right. \;.
\label{eq:nuflux2}
\end{equation}
We refer to the models based on Eq.(\ref{eq:nuflux2}) and 
Eq.(\ref{eq:nuflux1}) as models 1 and 2, respectively.
In Fig.~\ref{fig2a} we show the muon neutrino spectrum 
for our fiducial parameters in models 1 and 2.

\section{Appendix: Photomeson Interactions of Protons with External Photons in the Supranova Model}
\label{PWBphotons}

In this section, we consider the neutrino photoproduction on external 
photons in supranova GRBs (Guetta \& Granot 2002a). The external radiation 
field surrounding the final black hole is referred to as the pulsar wind 
bubble (PWB). 
A pulsar wind bubble 
is formed when the relativistic wind 
(consisting of relativistic particles and 
magnetic fields) that emanates from a pulsar is abruptly decelerated 
(typically, to a Newtonian velocity) in a strong relativistic shock, 
due to interaction with the ambient medium.

Fractions of the post-shock energy density go to the magnetic field, 
the electrons and the protons, respectively.
The electrons will lose energy through synchrotron emission and 
inverse-Compton (IC) scattering. In (Guetta \& Granot 2003) a deep
study of the characteristic features of the 
plerion emission has been carried out.
As shown in (Guetta \& Granot 2003), the electrons are
in the fast cooling regime for relevant values of $t_{\rm sd}$, and therefore
most of the emission takes place within a small radial interval just behind 
the wind termination shock.

The mechanism  for neutrino production through photomeson
interaction with the external photons dominates when the 
lifetime of the supramassive neutron star, 
$t_{\rm sd}\lesssim 0.2\;$yr (or $t_{\rm sd}\lesssim 2\;$yr for X-ray 
flashes). Neutrinos generated in this way have typical energies 
$\varepsilon_\nu\sim 10^{15}-10^{17}\,(10^{19})\;$eV ($10^{19} \;$eV for 
X-ray flashes). As in models 1 and 2, these neutrinos are emitted 
simultaneously with the prompt $\gamma$-ray (X-ray) emission. For even 
shorter lifetimes $t_{\rm sd}\lesssim 0.1\;$yr, the $\nu$'s would not be 
accompanied by a detectable GRB because the Thomson optical depth on the 
PWB is larger than unity.

As before, protons of energy $\varepsilon_p$ interact mostly with photons
that satisfy the $\Delta$-resonance condition,
$\varepsilon_{p,\Delta}=0.3\,{\rm GeV^2}/\varepsilon_\gamma$,
where, in this case, $\varepsilon_\gamma$ is the PWB photon energy.
The minimum photon energy for photomeson interactions corresponds to 
the maximum proton energy 
$\varepsilon_{\gamma,{\rm min}}=0.3\,{\rm GeV^2}/\varepsilon_{p,{\rm max}}\sim
3\times 10^{-3}(\varepsilon_{p,{\rm max}}/10^{20}{\rm eV})^{-1}\;$eV. 
For reasonable model parameters  ($t_{\rm sd}\gtrsim 0.026\;$yr) 
this energy exceeds self absorption frequency of the PWB spectrum 
(Guetta \& Granot 2003). Moreover, for $t_{\rm sd}\lesssim 12$ yr
the electrons are in fast cooling and emit synchrotron radiation.
Therefore the relevant part of the  spectrum consists of 
two power laws, 
$dn_\gamma/d\varepsilon_\gamma\propto\varepsilon_\gamma^{-3/2}$ for
$\varepsilon_\gamma<\varepsilon_{\gamma b}=h\nu_{bm}$ and
$dn_\gamma/d\varepsilon_\gamma\propto\varepsilon_\gamma^{-s/2-1}$
for $\varepsilon_\gamma>\varepsilon_{\gamma b}$, where
$\nu_{bm}\approx 1.6 \times 10^{15} t_{\rm sd,-1}^{-3/2}\;$\,Hz
is the peak frequency of the PWB spectral energy distribution
$(\nu F_\nu)$, $n_\gamma$ is the number density of photons and
$s\approx 2.2$ is the power law index of the PWB electrons.

The normalization factor of the target photon number density is
determined by equating the pulsar wind luminosity in pairs,
$\xi_e E_{\rm rot}/t_{\rm sd}$ ($\xi_{e}$ is the fraction of the
pulsar wind energy in the $e^\pm$ component and $E_{\rm rot}$ is
the total energy of the pulsar wind) to the total energy output in photons, 
which is $\sim U_{ph}4\pi R^2$ at $R\gg R_s=fR_b$, where $R_s$ 
is the radius of the pulsar wind termination shock, which is a 
factor, $f$, smaller than the PWB radius, $R_b$.
At $R\lesssim R_s$, which is relevant for
our case, the photons are roughly isotropic and $U_{ph}$ becomes roughly
constant, and assumes the value
$U_{ph}\approx\xi_{e}E_{\rm rot}/t_{\rm sd}2\pi(fR_b)^2 c$. 
As we mentioned above, the relevant target photons for photomeson interaction
with high energy protons are the synchrotron photons and the
fraction of the total photon energy that goes into the synchrotron 
component is $U_{syn}/U_{ph}=(1+Y_b)\approx\sqrt{\epsilon_{bB}/\epsilon_{be}}$,
where $\epsilon_{be}$ and $\epsilon_{bB}$
are the fractions of the PWB energy in the electrons and in magnetic field,
respectively.

This spectrum is consistent with the spectrum of known plerions like the
Crab, that can be well fit by emission from a power law distribution
with a power law index value $\sim 2.2$.
In our case, the only difference is that there is a fast
cooling synchrotron spectrum, rather than a slow cooling one. There are no
observations of the spectrum from very young plerions where there is a
fast cooling spectrum (as they are more rare).

This radiation will be typically hard to detect (for $t_{\rm sd}\lesssim 1$ yr
and $z\sim 1$), but might be detected for closer (though rarer) PWBs.

The proton energy satisfying the $\Delta$-resonance condition with PWB photons of energy $\varepsilon_{\gamma b}$, is
\begin{equation}
\varepsilon_{pb}=4.4\times 10^{16}{\xi_{e/3}^2\beta_{b,-1}^{3/2}
t_{\rm sd,-1}^{3/2}\over\eta_{2/3}^{5/2}\epsilon_{be/3}^{2}
\epsilon_{bB,-3}^{1/2}E_{53}^{1/2}\gamma_{w,4.5}^{2}}\;{\rm eV}\ ,
\end{equation}
where $t_{\rm sd,-1}=t_{\rm sd}/0.1\;$yr, $\beta_{b-1}=\beta_{b}/0.1$ is the
velocity of the SNR shell (in units of $c$), 
$E_{53}=E_{\rm{rot}}/10^{53}\;$erg,
$\gamma_{w,4.5}=\gamma_{w}/10^{4.5}$ is the Lorentz
factor of the pulsar wind, $\xi_{e}=\xi_{e/3}/3$,
$\epsilon_{be}=\epsilon_{be/3}/3$, $\epsilon_{bB}=10^{-3}\epsilon_{bB,-3}$,
and $\eta=(2/3)\eta_{2/3}$ is the fraction of the wind
energy that remains in the PWB. The latter is the fraction that goes into the
proton component and is, unlike the electron component, not radiated away. The corresponding neutrino energy is
$\varepsilon_{\nu b}\approx\varepsilon_{pb}/20
\sim 2\times 10^{15}t_{\rm sd,-1}^{3/2}\;{\rm eV}$.
For $t_{\rm sd}\sim 0.1\;$yr, this energy is similar to those obtained in interactions with GRB photons in the previous
section. Photon emission is only detected in coincidence with these neutrinos if the Thompson optical depth is $\lesssim 1$, which is the case for $t_{\rm sd}\sim 0.1\;$yr and a clumpy SNR (Guetta \& Granot 2003; Inoue, Guetta \& Pacini 2003). In the case of a uniform shell the condition is $t_{\rm sd}\gtrsim 0.4\;$yr, corresponding to $\varepsilon_{\nu b}\gtrsim 2\times 10^{16}\;$eV.

As before, the internal shocks occur over a distance $R=2\Gamma^2 c t_v$.
Thus, the optical depth for photo-pion production by protons of energy $\varepsilon_p$, is
\begin{equation}\label{tau_pg_GRB}
\tau_{p\gamma}= 
\sigma_{p\gamma}\varepsilon_\gamma{dn_\gamma\over d\varepsilon_\gamma}R=
{1.0\xi_{e/3}^3E_{53}^{1/2}
\Gamma_{2.5}^2 t_{v,-2}(\varepsilon_p/\varepsilon_{pb})^\beta\over
f_{1/3}^{2}\eta_{2/3}^{5/2}\epsilon_{be/3}^{5/2}
\gamma_{p,4.5}^{2}\beta_{b,-1}^{1/2}t_{\rm sd,-1}^{3/2}}
\end{equation}
where $\Gamma_{2.5}=\Gamma/10^{2.5}$, $t_{v,-2}=t_v/10^{-2}\;$s,
$\sigma_{p\gamma}\approx 5\times 10^{-28}\;{\rm cm^2}$,
$\varepsilon_\gamma=0.3\,{\rm GeV^2}/\varepsilon_p$,
$f=f_{1/3}/3$ and $\beta$ is the spectral slope of the
seed PWB synchrotron photons with $\beta=s/2$ ($1/2$) for
$\varepsilon_p<\varepsilon_{pb}$ ($\varepsilon_p>\varepsilon_{pb}$).
The fraction of the proton energy that is lost to pion
production is given by
\begin{equation}\label{f_pg_GRB}
f_{p\pi}(\varepsilon_p)\approx 1-\exp\left[-\tau_{p\gamma}(\varepsilon_p)/5\right]
\approx\min\left[1,\tau_{p\gamma}(\varepsilon_p)/5\right]\ .
\end{equation}
The factor of 5, as before, takes into account that the proton
loses $\sim 0.2$ of its energy in a single interaction.
We denote $\varepsilon_{p}$ for which $f_{p\pi}(\varepsilon_{p})\approx 1$
by $\varepsilon_{p\tau18}=\varepsilon_{p\tau}/10^{18}\;$eV
[i.e. $\tau_{p\gamma}(\varepsilon_{p\tau})\equiv 5$], and obtain
\begin{equation}\label{e_tau}
\varepsilon_{p\tau18}=\left\{\begin{array}{lll}
{0.20f_{1/3}^{20/11}\eta_{2/3}^{25/11}
\epsilon_{be/3}^{3/11}\beta_{b,-1}^{43/22}t_{\rm sd,-1}^{63/22}
\over\xi_{e/3}^{8/11}\epsilon_{bB,-3}^{1/2}E_{53}^{21/22}
\gamma_{p,4.5}^{2/11}\Gamma_{2.5}^{20/11}t_{v,-2}^{10/11}}&
{\varepsilon_{p\tau}\over\varepsilon_{pb}}<1 \\
{1.2f_{1/3}^{4}\eta_{2/3}^{5/2}\epsilon_{be/3}^{3}
\gamma_{p,4.5}^{2}\beta_{b,-1}^{5/2}t_{\rm sd,-1}^{9/2}
\over \xi_{e/3}^{4}\epsilon_{bB,-3}^{1/2}E_{53}^{3/2}
\Gamma_{2.5}^{4}t_{v,-2}^{2}}&
{\varepsilon_{p\tau}\over\varepsilon_{pb}}>1 \end{array}\right. \ .
\end{equation}
As before, the decay of charged pions created in interactions between PWB photons and
GRB protons, produces high energy neutrinos, $\pi^+\rightarrow \mu^+ +\nu_{\mu}
\rightarrow e^+ +\nu_e +\bar{\nu}_{\mu} +\nu_{\mu}$, where each neutrino
receives $\sim 5\%$ of the proton energy.

The total energy of the protons accelerated in the internal shocks is expected to
be similar to the $\gamma$-ray energy produced in the GRB (Waxman 1995).
This implies a $\nu_\mu$ fluence,
\begin{equation}\label{F_nu_GRB}
f_{\nu_\mu}=f_{0}f_{p\nu}\ \ \ , \ \ \ f_{0}={E_{\gamma,iso}\over 32\pi d_L^2}
=1\times 10^{-5}{E_{\gamma,53}\over d_{L28}^{2}}\;{\rm erg\over cm^{2}}\ ,
\end{equation}
$$
f_{p\nu}={\int d\varepsilon_p(dN_p/d\varepsilon_p)\varepsilon_p f_{p\nu}(\varepsilon_p)
\over\int d\varepsilon_p(dN_p/d\varepsilon_p)\varepsilon_p}\ ,
$$
where $E_{\gamma,53}=E_{\gamma,\rm{iso}}/10^{53}\;$erg is the
isotropic equivalent energy in $\gamma$-rays,
$f_{p\nu}(\varepsilon_p)=f_{p\pi}(\varepsilon_p)f_{\pi\nu}(\varepsilon_p)$,
while $f_{p\pi}(\varepsilon_p)$ is given in Eq.(\ref{f_pg_GRB}) and
$f_{\pi\nu}(\varepsilon_p)$ is the fraction of the original pion energy,
$\varepsilon_\pi\approx 0.2\varepsilon_p$,
that remains after decay.

The pions may lose energy via synchrotron or inverse Compton (IC) emission. If these energy
losses are significant, then the energy of the neutrinos will be reduced as well.  Following arguments already presented in appendix A, we find that
$f_{\pi\nu}(\varepsilon_p)\approx 1-\exp(-t'_{\rm rad}/\tau'_\pi)
\approx \min(1,t'_{\rm rad}/\tau'_\pi)$, where
$\tau'_\pi\approx 2.6\times 10^{-8}\varepsilon'_\pi/(m_\pi c^2)\;$s
is the lifetime of the pion, and
$t'_{\rm rad}=(t_{\rm syn}^{\prime -1}+t_{\rm IC}^{\prime -1})^{-1}\approx
\min(t'_{\rm syn},t'_{\rm IC})$ is the time for radiative losses due to both
synchrotron and IC losses. The time, $t'_{\rm syn}$, is given by Eq.(\ref{t_sync}) and
\begin{equation}\label{t_rad}
t'_{\rm IC}={3m_\pi^4c^3\over 4\sigma_T m_e^2 \varepsilon_\pi U'_\gamma
(\varepsilon_\pi)}\ ,
\end{equation}
where $U'_\gamma(\varepsilon_\pi)$ is the
energy density of photons below the Klein-Nishina limit,
$\varepsilon_{\gamma,KN}=(m_\pi c^2)^2/\varepsilon_\pi$.
IC losses due to scattering of the GRB photons were shown to be
unimportant (Waxman \& Bahcall 1997). We therefore only consider the IC losses from the
upscattering of the external PWB photons, and find
\begin{equation}\label{t_IC}
{t'_{\rm IC}/\tau'_\pi}= 0.67f_{1/3}^2\xi_{e/3}^{-1}
\epsilon_{be/3}^{1/2}\epsilon_{bB,-3}^{-1/2}E_{53}^{-1}
\beta_{b,-1}^{2}t_{\rm sd,-1}^3\varepsilon_{\pi18}^{-2}\,\, ,\,
\end{equation}
where $\epsilon_e$ and $\epsilon_B$
are the equipartition parameters of the GRB, and
$\varepsilon_{\pi18}=\varepsilon_{\pi}/10^{18}\;$eV.
The radiative losses become important for $t'_{\rm rad}<\tau'_\pi$,
which corresponds to $\varepsilon_{p}>\varepsilon_{ps}=
\min(\varepsilon_{ps}^{\rm syn},\varepsilon_{ps}^{\rm IC})
\approx 5\varepsilon_{\pi s}
\approx 20\varepsilon_{\nu s}$, where
\begin{eqnarray}\label{e_syn}
\varepsilon_{ps18}^{\rm syn}= 2.3\,\epsilon_e^{1/2}
\epsilon_B^{-1/2}L_{52}^{-1/2}\Gamma_{2.5}^4 t_{v,-2}\ ,
\quad\quad\quad\quad\quad\ \ \, \\ 
\label{e_IC}
\varepsilon_{ps18}^{\rm IC}= 4.1f_{1/3}\xi_{e/3}^{-1/2}
\epsilon_{be/3}^{1/4}\epsilon_{bB,-3}^{-1/4}E_{53}^{-1/2}
\beta_{b,-1}t_{\rm sd,-1}^{3/2}\ .\
\end{eqnarray}

The protons may also lose energy via $p-\gamma$ interactions with 
the GRB photons (Waxman \& Bahcall 1997). However $\tau_{p\gamma}$
for this process is typically $< 1$, so that it does not have 
a large effect on $p\gamma$ interactions with the PWB photons, on which
we focus.

Because the lifetime of the muons is $\sim\,$100 times longer than
that of the pions,
they experience significant radiative losses at an energy of
$\varepsilon_{\mu s}\sim\varepsilon_{\pi s}/10\approx\varepsilon_{ps}/50$.
This causes a reduction of up to a factor of $3$ in the total neutrino flux
in the range $\sim (0.1-1)\varepsilon_{\nu s}$,
since only $\nu_\mu$ that are produced directly in $\pi^+$ decay contribute
significantly to the neutrino flux.
Note that since both ratios in Eqs. (\ref{t_syn}) and (\ref{t_IC})
scale as $\varepsilon_{\pi}^{-2}$, we always have
$t'_{\rm rad}/\tau'_\pi\propto\varepsilon_{\pi}^{-2}$, and therefore,
the spectrum steepens by a factor of
$(\varepsilon_\nu/\varepsilon_{\nu s})^{-2}$ for
$\varepsilon_\nu>\varepsilon_{\nu s}$. This is evident because $f_{\pi\nu}(\varepsilon_p\approx 5\varepsilon_\pi)
\approx\min(1,t'_{\rm syn}/\tau'_\pi,t'_{\rm IC}/\tau'_\pi)$.

\begin{figure}
\includegraphics[width=14cm]{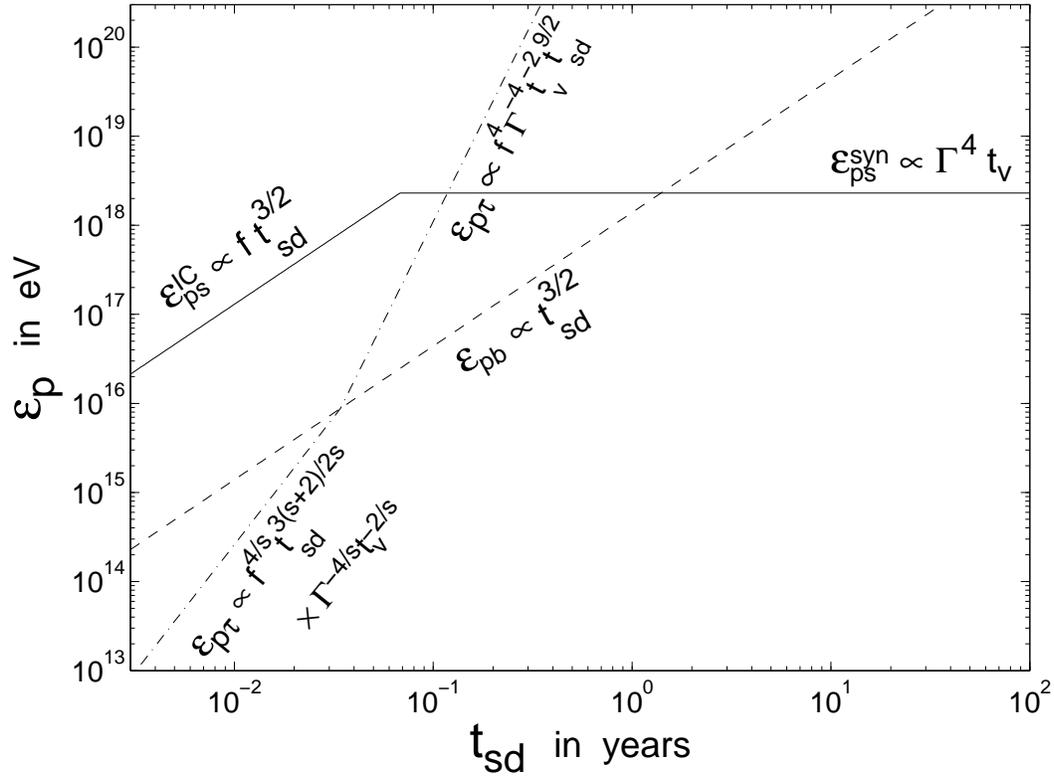}
\caption{\label{fig1}
The proton energies that correspond to break energies in the
neutrino spectrum, $\varepsilon_\nu\approx\varepsilon_p/20$,
as a function of $t_{\rm sd}$.
This figure is taken from Guetta \& Granot 2002a}
\end{figure}

In Figure \ref{fig1} we show the proton energies that correspond to the
neutrino break energies $\varepsilon_{\nu s}$, $\varepsilon_{\nu b}$ and
$\varepsilon_{\nu\tau}\approx\varepsilon_{p\tau}/20$, as a function of $t_{\rm sd}$.
From Eq. (\ref{tau_pg_GRB}), (\ref{e_syn}) and (\ref{e_IC}), we conclude that $\tau_{p\gamma}\propto\Gamma^2 t_{\rm sd}^{-3/2}$, while
$\varepsilon_{ps}^{\rm syn}\propto\Gamma^4$ and
$\varepsilon_{ps}^{\rm IC}\propto t_{\rm sd}^{3/2}$.
For a fixed value of $t_{\rm sd,-1}=1$,  $\Gamma_{2.5}\gtrsim 1$ implies
$\tau_{p\gamma}\gtrsim 1$ and
$\varepsilon_{ps}=\varepsilon_{ps}^{\rm IC}={\rm constant}\sim 10^{18}\;$eV,
while $\Gamma_{2.5}\lesssim 1$ implies
$\varepsilon_{ps}=\varepsilon_{ps}^{\rm syn}\propto\Gamma^4$.
This implies increased neutrino emission reaching higher energies
for larger values of $\Gamma$. Since large $\Gamma$ implies lower synchrotron frequency for the prompt GRB, this may be relevant to X-ray flashes, assuming that they are indeed GRBs with relatively large Lorentz factors and/or
a large variability time, $t_v$ (Guetta, Spada \& Waxman 2001b).

As can be seen from Figure \ref{fig1}, depending on the relevant parameters,
there are four different orderings of these break energies:
(i) $\varepsilon_{\nu\tau}<\varepsilon_{\nu b}<\varepsilon_{\nu s}$,
(ii) $\varepsilon_{\nu b}<\varepsilon_{\nu\tau}<\varepsilon_{\nu s}$,
(iii) $\varepsilon_{\nu b}<\varepsilon_{\nu s}<\varepsilon_{\nu\tau}$ and (iv) $\varepsilon_{\nu s}<\varepsilon_{\nu b}<\varepsilon_{\nu\tau}$.
Each of these results in a different spectrum consisting of 3 or 4 power laws.
Analytically these spectra are:

\begin{equation}\label{spectrum1}
{\varepsilon_\nu^2{dN_\nu\over d\varepsilon_\nu}(i)/f_0\;\over
{(4+s)/4s+\ln(\varepsilon_{\nu s}/\varepsilon_{\nu\tau})\over
\left[(4+s_/2s+\ln(\varepsilon_{\nu s}/\varepsilon_{\nu\tau})\right]^{2}}}
=
\left\{\begin{array}{lll}
(\varepsilon_{\nu}/ \varepsilon_{\nu\tau})^{s/2}
& \ \ \varepsilon_{\nu}<\varepsilon_{\nu\tau}\\
1 & \ \ \varepsilon_{\nu\tau}<\varepsilon_{\nu}<\varepsilon_{\nu s}\\
(\varepsilon_{\nu}/\varepsilon_{\nu s})^{-2} 
& \ \  \varepsilon_{\nu}>\varepsilon_{\nu s}
\end{array}\right. \ ,
\end{equation}
\begin{equation}\label{spectrum2}
{\varepsilon_\nu^2{dN_\nu\over d\varepsilon_\nu}(ii)/f_0\;\over
{5/4+\ln(\varepsilon_{\nu s}/\varepsilon_{\nu\tau})\over
\left[5/2+\ln(\varepsilon_{\nu s}/\varepsilon_{\nu\tau})\right]^{2}}}
=
\left\{\begin{array}{lll}\sqrt{\varepsilon_{\nu b}\over\varepsilon_{\nu\tau}}
\left({\varepsilon_{\nu}\over\varepsilon_{\nu b}}\right)^{s\over 2}
& \ \ \varepsilon_{\nu}<\varepsilon_{\nu b}\\
(\varepsilon_{\nu}/\varepsilon_{\nu\tau})^{1/2}
& \ \ \varepsilon_{\nu b}<\varepsilon_{\nu}<\varepsilon_{\nu\tau}\\
1 & \ \  \varepsilon_{\nu\tau}<\varepsilon_{\nu}<\varepsilon_{\nu s}\\
(\varepsilon_{\nu}/\varepsilon_{\nu s})^{-2}
& \ \  \varepsilon_{\nu}>\varepsilon_{\nu s}\end{array}\right. \ ,
\end{equation}

\begin{equation}\label{spectrum3}
{\varepsilon_\nu^2{dN_\nu\over d\varepsilon_\nu}(iii)/f_0
\over
{3\over 16}\sqrt{\varepsilon_{\nu s}\over\varepsilon_{\nu\tau}}}=
\left\{\begin{array}{lll}
\left({\varepsilon_{\nu b}\over\varepsilon_{\nu s}}\right)^{1\over 2}
\left({\varepsilon_{\nu}\over\varepsilon_{\nu b}}\right)^{s\over 2}
& \ \ \varepsilon_{\nu}<\varepsilon_{\nu b}\\
(\varepsilon_{\nu}/\varepsilon_{\nu s})^{1/2}
& \ \ \varepsilon_{\nu b}<\varepsilon_{\nu}<\varepsilon_{\nu s}\\
(\varepsilon_{\nu}/\varepsilon_{\nu s})^{-3/2}
& \ \  \varepsilon_{\nu s}<\varepsilon_{\nu}<\varepsilon_{\nu\tau}\\
\left({\varepsilon_{\nu s}\over\varepsilon_{\nu\tau}}\right)^{3\over 2}
\left({\varepsilon_{\nu\tau}\over\varepsilon_{\nu}}\right)^2
& \ \  \varepsilon_{\nu}>\varepsilon_{\nu\tau}\end{array}\right. \ ,
\end{equation}

\begin{equation}\label{spectrum4}
{\varepsilon_\nu^2{dN_\nu\over d\varepsilon_\nu}(iv)/f_0
\over [s(4-s)/16](\varepsilon_{\nu s}/\varepsilon_{\nu b})^{s/2}
\sqrt{\varepsilon_{\nu b}/\varepsilon_{\nu\tau}}}=
\left\{\begin{array}{lll}
(\varepsilon_{\nu}/\varepsilon_{\nu s})^{s/2} 
& \ \ \varepsilon_{\nu}<\varepsilon_{\nu s}\\
(\varepsilon_{\nu}/\varepsilon_{\nu s})^{(s-4)/2} 
& \ \ \varepsilon_{\nu s}<\varepsilon_{\nu}<\varepsilon_{\nu b}\\
\left({\varepsilon_{\nu b}\over\varepsilon_{\nu s}}\right)^{(s-4)/2}
\left({\varepsilon_{\nu}\over\varepsilon_{\nu b}}\right)^{-3/2}
& \ \  \varepsilon_{\nu b}<\varepsilon_{\nu}<\varepsilon_{\nu\tau}\\
\left({\varepsilon_{\nu b}\over\varepsilon_{\nu s}}\right)^{s-4\over 2}
\left({\varepsilon_{\nu b}\over\varepsilon_{\nu\tau}}\right)^{3\over 2}
\left({\varepsilon_{\nu\tau}\over\varepsilon_{\nu}}\right)^{2}
& \ \  \varepsilon_{\nu}>\varepsilon_{\nu\tau}
\end{array}\right. \ .
\end{equation}
For our analysis we consider two characteristic values
of $t_{\rm sd}$, 0.07 yr and 0.4 yr, and refer to the models as 3 and 4,
respectively. In the case of Model 3, the GRB is seen
only if the shell is sufficiently clumpy while in model 4 the GRB should always be detectable.

\begin{figure}
\includegraphics[width=14cm]{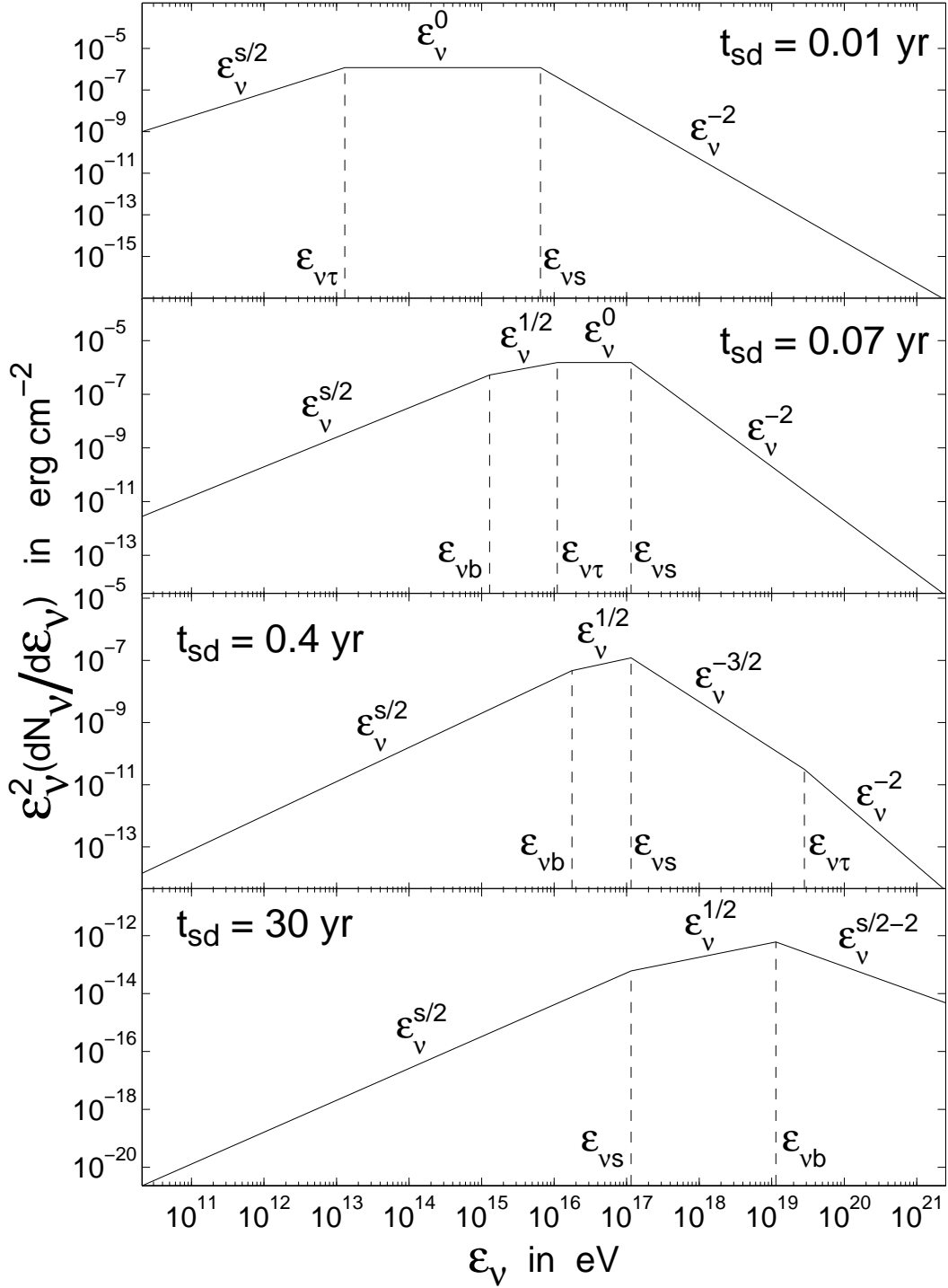}
\caption{\label{fig2}
The muon neutrino spectrum, $\varepsilon_\nu^2(dN_\nu/d\varepsilon_\nu)$,
for our fiducial parameters in models 3 and 4 (interactions with PWB photons). 
Four choices of $t_{\rm sd}=0.01,\, 0.07,\, 0.4,\, 30\;$yr are used,
which correspond to the 4 different orderings of the break energies.
This figure is taken from Guetta \& Granot 2002a}
\end{figure}

The muon neutrino spectrum is shown in Figure \ref{fig2} for our fiducial
parameters and $t_{\rm sd}=0.01,\, 0.07,\, 0.4,\, 30\;$yr.
The spectrum of the other neutrino flavors
is the same.

\section{Appendix: Calculations of Event Rates}

A compilation of the probability that a GRB neutrino is actually detected as a muon, a tau or a shower by an underground detector is shown in Fig.\,4 as a function of the neutrino energy. These are required to convert neutrino spectra from GRBs to event rates. We present in this appendix the formalism for doing this conversion.

\subsection{Showers}

The number of shower events in an underground detector from a 
neutrino flux $\Phi_{\nu_i}$ produced by a single GRB with duration
$T$ is given by
\begin{eqnarray}
N_{\rm sh} &=& \sum _{i,j} 2\pi A T  
\int dE_{\nu_i} \frac{d\Phi_{\nu_i}}{dE_{\nu_i}}(E_{\nu_i}) \,
P_{\rm surv}(E_{\nu_i},\theta_z) \nonumber \\
&& \times\int_{y^{i,j}_{\rm min}}^{y^{i,j}_{\rm max}} dy 
\frac{1}{\sigma^j(E_{\nu_i})} \,
\frac{d\sigma^j}{dy}(E_{\nu_i},y) \, P_{\rm int}(E_{\nu_i},y,\theta_z) \ ,
\label{rate}
\end{eqnarray}
where $\theta_z$ is the zenith angle ($\theta_z = 0$ is vertically
downward). The sum is over neutrino (and anti-neutrino) flavors
$i=e, \mu, \tau$ and interactions $j= {\rm CC}$ (charged current) and
NC (neutral current).  $A$ is the detector's cross sectional area with
respect to the $\nu$ flux, and
$d\Phi_{\nu_i}/dE_{\nu_i}$ is the differential neutrino flux that
reaches the Earth.  For $i=\tau$, \ref{rate} is modified to include
the effects of regeneration of neutrinos propagating through the Earth, as will be discussed further on.

$P_{\rm surv}$ is the probability that a neutrino reaches the detector, i.e. is not absorbed by the Earth.  It is given by
\begin{equation} 
P_{\rm surv} \equiv 
\exp [-X(\theta_z)\sigma^{\rm tot}(E_{\nu_i}) N_A] \ ,
\label{survival}
\end{equation}
where $N_A \simeq 6.022 \times 10^{23}{\rm g}^{-1}$, and the total
neutrino interaction cross section is
\begin{equation}
\sigma^{\rm tot} = \sigma^{\rm CC} + \sigma^{\rm NC} \ .
\end{equation}
This is somewhat conservative because it neglects the possibility of a
neutrino interacting via a NC interaction and subsequently creating a shower in the detector.  $X(\theta_z)$ is the column density of
material a neutrino with zenith angle $\theta_z$ must traverse to reach the detector. It depends on the depth of the detector and is given by
\begin{equation}
X(\theta_z)=\int_{\theta_z} \rho(r(\theta_z,l)) \, dl \ ,
\end{equation} 
the path length along direction $\theta_z$ weighted by the Earth's
density $\rho$ at distance $r$ from the Earth's center.  For the
Earth's density profile we adopt the piecewise continuous density
function $\rho(r)$ of the Preliminary Earth Model (Dziewonski 1989).

$P_{\rm int}$ is the probability that the neutrino interacts in the
detector. It is given by
\begin{equation}
P_{\rm int} = 1 - \exp \left[ - \frac{L}{L^j (E_{\nu_i})}
\right] \ ,
\end{equation}
where, for showers, $L$ is the linear dimension of the detector, and
$L^j (E_{\nu_i})$ is the mean free path for neutrino
interaction of type $j$.  For realistic detectors, $L \ll L^j
(E_{\nu_i})$, and so $P_{\rm int} \approx L/L^j (E_{\nu_i})$.
To an excellent approximation the event rate scales
linearly with detector volume $V=AL$. 

The inelasticity parameter $y$ is the fraction of the initial neutrino energy carried by the hadronic shower (rather than the primary lepton).  The limits of integration depend on the type of interaction and on the neutrino flavor.  For NC
$\nu_e$ interactions and all $\nu_\mu$ and $\nu_\tau$ interactions,
$y_{\rm max}=1$ and $y_{\rm min}=E^{\rm thr}_{\rm sh}/E_\nu$, where
$E^{\rm thr}_{\rm sh}$ is the threshold energy for shower
detection. For CC $\nu_e$ interactions, the outgoing electron also
showers, therefore $y_{\rm max}=1$ and $y_{\rm min}=0$.

\subsection{Muons}

Energetic muons are produced in $\nu_\mu$ CC
interactions.  For a muon to be detected, it must reach the detector
with an energy above its threshold $E^{\rm thr}_\mu$.  The expression of
Eq.(\ref{rate}) then also describes the number of muon events after the replacement
\begin{equation}
P_{\rm int} = 1-\exp\left[-\frac{R_\mu(E_\mu,\theta_z)}
{L^{\rm CC}(E_{\nu_\mu})} \right] \ ,
\end{equation}
where $R_\mu$ is the range of a muon with initial energy $E_\mu =
(1-y) E_{\nu_\mu}$ and final energy $E^{\rm thr}_\mu$. We will assume that 
muons lose energy continuously according to
\begin{equation}
\frac{dE}{dX}=-\alpha - \beta E \ ,
\end{equation}
where $\alpha=2.0~{\rm MeV} {\rm cm}^2/{\rm g}$ and $\beta=4.2 \times
10^{-6}~{\rm cm}^2/{\rm g}$ (Dutta et al. 2001). The muon range is then
\begin{equation}  
R_\mu = \frac{1}{\beta} \ln \left[ 
\frac{\alpha + \beta E_\mu}{\alpha + \beta E^{\rm thr}_\mu} \right]
\ .
\label{murange}
\end{equation}
In this case, $y_{\rm max} = 1 - E^{\rm thr}_\mu/E_\nu$ and $y_{\rm
min}=0$.

The event rate for muons is enhanced by the possibility that muons reach the detector, even if produced in neutrino interactions kilometers from its location.  Note,
however, that this enhancement (i.e. $R_\mu$) 
is $\theta_z$-dependent: for nearly
vertical down-going paths, the path length of the muon is limited by
the amount of matter above the detector, not by the muon's range.
This is taken into account in the simulations.
Fig.\ref{fig5} shows
the probability of detecting a muon generated by a muon neutrino
as a function of the incidence zenith angle. The figure illustrates
the effect of the limited amount of matter above the detector,
as well as the neutrino absorption in the Earth.
Absorption is not important at $E_\nu=1$ TeV, and the muon range at this
energy (for a muon energy threshold of 500 GeV as we adopted in the
figure) is not
limited by the amount of matter above the detector. As
a consequence the probability at 1 TeV is weakly dependent on zenith angle.
Absorption in the Earth starts to be important for upgoing neutrinos
of energy above $\sim 100$ TeV - 1 PeV, and restricts the neutrino
observation to the horizontal and downward directions at extremely high
energy (EeV range) as can be seen in the figure. The large muon range at PeV
and EeV energies, much larger than the depth at which the detector is
located (1.8 km vertical depth), reduces considerably the detection 
probability of downgoing neutrinos.

The largest background to a GRB signal consists of muons from atmospheric neutrinos. However, for GRB observations, the time and angular windows are very small and this background can be easily controlled, as we will illustrate. Following (Dermer \& Atoyan 2003), the number of background events is approximately given by
\begin{equation}
N_{\rm{bg}} \simeq A \int d\Omega \int dt \int^{\infty}_{E_{\rm{min}}} dE_{\nu} \frac{dN_{\nu}}{dE_{\nu}} P_{\nu \rightarrow \mu}(E_{\nu})   
\label{background}
\end{equation}
where $\Omega$ and $t$ are the solid angle and time considered, respectively, and $P_{\nu \rightarrow \mu}(E_{\nu})$ is the probability of a muon neutrino generating a muon in the detector volume. IceCube is designed to have angular resolution smaller than 1 degree at the relevant energies. We consider a 1 degree cone for the solid angle calculation. Considering a long burst of duration $\sim 100$ seconds, and using the following approximate atmospheric neutrino spectrum    
\begin{equation}
\frac{dN_{\nu}}{dE_{\nu}}
=
\left\{\begin{array}{lll}
5 \times 10^{-18} E_{\nu}^{-4} \,\rm{GeV}\,\rm{cm}^{-2}\,\rm{s}^{-1}\,\rm{sr}^{-1}
& \ \ E_{\nu}>10^5 \rm{GeV}\\
1.6 \times 10^{-16} E_{\nu}^{-3.7}\,\rm{GeV}\,\rm{cm}^{-2}\,\rm{s}^{-1}\,\rm{sr}^{-1}
& \ \ \ E_{\nu}<10^5 \rm{GeV}\\
\end{array}\right. \ ,
\label{background2}
\end{equation}
we find that for a natural threshold (minimum energy) of $\sim 100\,$GeV, we expect $\sim .003$ background events per burst. More practically, an energy threshold in the range of 1-10 TeV could be imposed which would reduce this background by an addition factor of 50 to 2500, respectively. For a naive illustration, consider one years of observation, with 1000 bursts, each of duration of 10 seconds and a 1 TeV energy threshold imposed. For such a example, less than 0.01 total background events are predicted. 

\subsection{Taus}

Taus are produced only by CC $\nu_\tau$ interactions. This process
differs significantly from the muon case because tau neutrinos are
regenerated by the production and subsequent tau decay through $\nu_\tau \to \tau \to
\nu_\tau$ (Halzen \& Saltzberg 1998). As a result, for tau neutrinos, CC and
NC interactions in the Earth do not deplete the $\nu_\tau$ flux, they only reduce the neutrino energy down to a value where, eventually, the Earth becomes transparent. We implement this important effect using a
dedicated simulation that determines $\overline{E}_{\nu_\tau}
(E_{\nu_\tau}, \theta_z)$, the average energy of the $\nu_\tau$ when reaching the detector. It depends on the initial energy
$E_{\nu_\tau}$ and zenith angle $\theta_z$.  The tau event rate is
then given by
\begin{eqnarray}
N_{\tau} &=& 2\pi A T 
\int dE_{\nu_\tau} \frac{d\Phi_{\nu_\tau}}{dE_{\nu_\tau}}(E_{\nu_\tau})
\int_{y_{\rm min}}^{y_{\rm max}} dy
\frac{1}{\sigma^{\rm CC}(\overline{E}_{\nu_\tau})}
\frac{d\sigma^{\rm CC}}{dy}(\overline{E}_{\nu_\tau},y)
\nonumber \\
&& \times P_{\rm int}
\Theta ((1-y)\overline{E}_{\nu_{\tau}} - E^{\rm thr}_\tau) \ , 
\label{taurate}
\end{eqnarray}
$P_{\rm int}$ depends on the geometry of the neutrino tau induced event.
For events consisting on a minimum-ionizing track going through the detector
it is given by:
\begin{equation}
P_{\rm int}=\left[1- \exp \left(-\frac{R_\tau((1-y)
\overline{E}_{\nu_\tau})}
{L^{\rm CC}(\overline{E}_{\nu_\tau})}\right)\right]
\label{eq:Pint_tau}
\end{equation}
where $R_\tau((1-y) \overline{E}_{\nu_\tau})$ is the range of the
produced tau evaluated at the energy of the tau after
regeneration.  $R_\tau$ is given by Eq.(\ref{murange}) with
$\beta=3.6 \times 10^{-7}~{\rm cm}^{-2}/{\rm g}$ (Dutta et al. 2001).  The last
factor takes into account the requirement that the tau track be long
enough to be identified in the detector.  We require $E^{\rm thr}_\tau
\simeq 2.5 \times 10^6~\;$GeV so that the tau decay length is larger than the 125\, m  string spacing in IceCube.  It is not clear with what efficiency through-going tau events can be separated from low energy muons.  

\begin{figure}
\includegraphics[width=14cm]{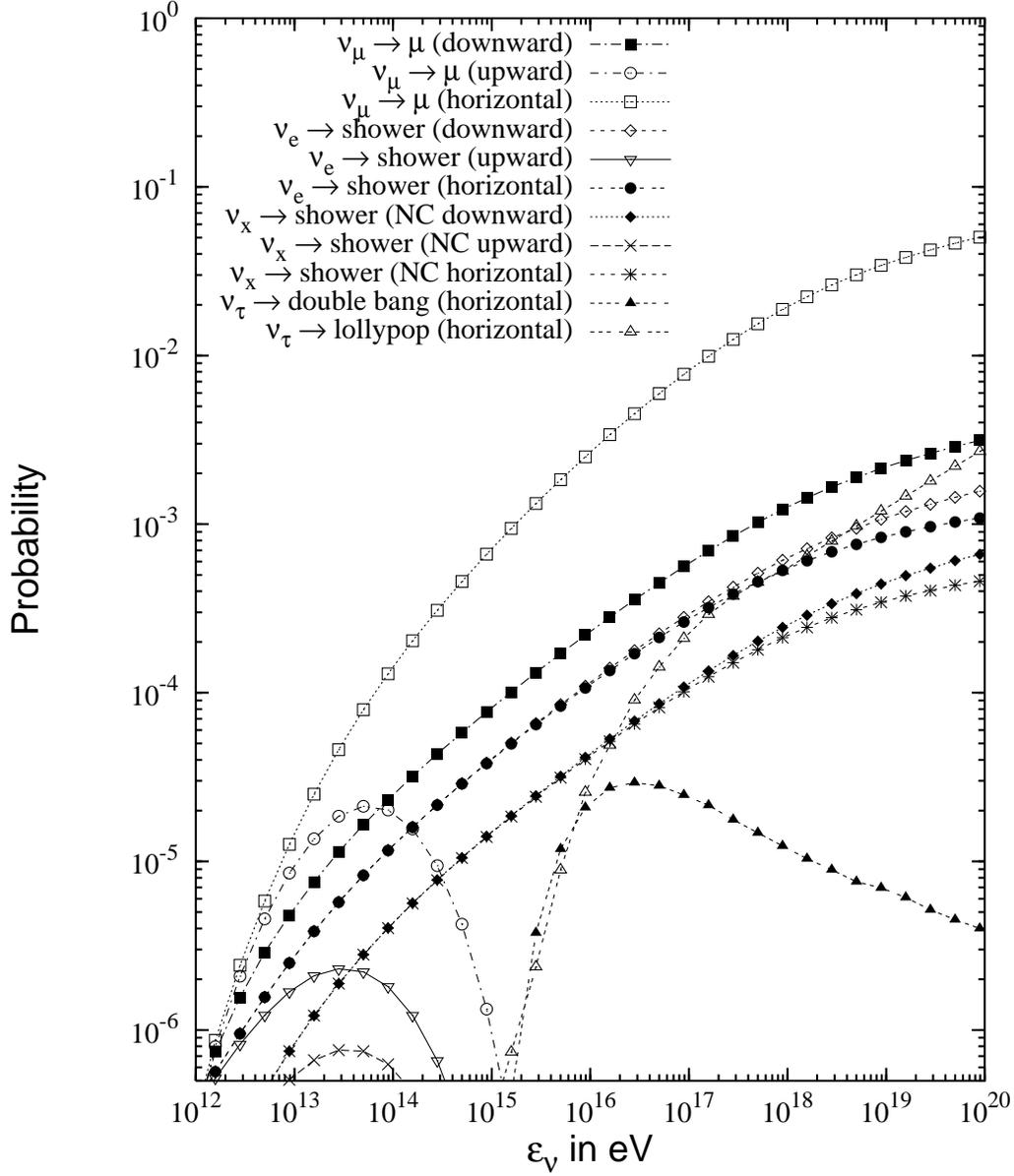}
\caption{\label{fig4}
The probabilities of a neutrino generating various types of events when 
traveling through the effective area of a neutrino telescope. Curves are 
shown for various choices of zenith angle, which reflects enhancements 
due to long muon range and the effect of absorption in the Earth.}
\end{figure}

\begin{figure}
\includegraphics[width=14cm]{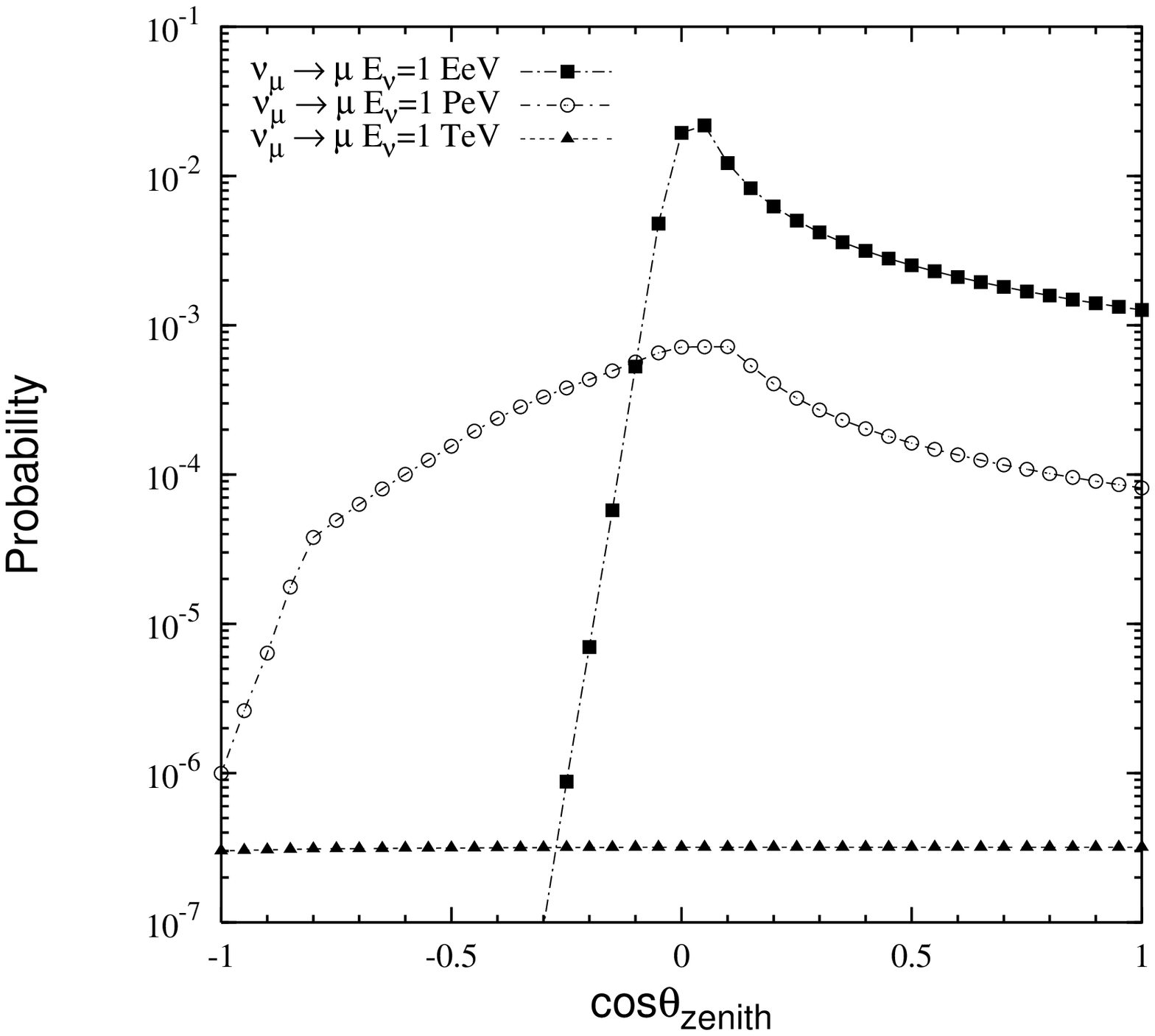}
\caption{\label{fig5}
The probability of a muon neutrino generating a detectable muon as a function
of zenith angle. The curves show the probabilities for three representative
neutrino energies: $10^{12}$ eV, $10^{15}$ eV and $10^{18}$ eV.}
\end{figure}

Those tau events that include one shower (lollipop
events) or two showers (double bang events) inside the detector volume
will be identifiable. For these cases we use $P_{\rm int}$ obtained by
a dedicated simulation that determines the probabilities for double 
bang and lollipop geometries shown in Fig.(\ref{fig4}). 
The rate of downgoing lollipop events
in a km$^3$ neutrino telescope is expected to be of the order of the
rate of down-going shower events, probably slightly smaller.  Double
bang events will be mostly observed for neutrino energies in a limited
range between roughly 10 and 100 PeV.

Finally, those events in which the tau decays into muons 
that reach the detector are counted as muon events. $P_{\rm int}$
is given by Eq.(\ref{eq:Pint_tau}) where we use as range the 
sum of the range of the tau evaluated at the regeneration energy 
and the range of the muon at the energy that carries in the decay of the 
tau. 

As with muons, at very high energies taus can travel several
kilometers before decaying or suffering significant energy loss.  The
enhancement to tau event rates from this effect is
$\theta_z$-dependent as discussed above for muons.

\newpage

\clearpage


\begin{thebibliography}{99}

\bibitem{Alvarez-Muniz:2000st}
Alvarez-Mu\~niz, J., Halzen, F. \& Hooper D. W., 2000,
PRD, 62, 093015.

\bibitem{AMANDA}
Andres, E. et al., 2001, Nature 410, 441.

\bibitem{antares}
Aslanides, E., et al. for the ANTARES Collaboration, 1999, astro-ph/9907432.

\bibitem{hardtke}
Barouch, G \& Hardtke, R., talk given at the 
International Cosmic Ray Conference, 2001. 

\bibitem{dermer}
Dermer, C. \& Atoyan, A., 2003, submitted to PRL, astro-ph/0301030.

\bibitem{Dutta:2000hh}
Dutta, S. I., Reno, M. H., Sarcevic, I. \& Seckel, D.,
2001, PRD, 63, 094020.

\bibitem{preliminary}
Dziewonski, A. ``Earth Structure, Global,'' in {\em The
Encyclopedia of Solid Earth Geophysics}, edited by D.~E.~James (Van
Nostrand Reinhold, New York, 1989), p.331.

\bibitem{E89}
Eichler, D., et al., 1989, Nature, 340, 126

\bibitem{feni}
Fenimore, E. E. \& Ramirez-Ruiz, E., 2000, in press on ApJ, astro-ph/0004176.

\bibitem{Fishman} 
Fishman, G. J. \& Meegan, C. A., 1995, ARA\&A {\bf 33}, 415.

\bibitem{Freedman}
Freedman, D., Waxman, E., 2001, ApJ, 547.

\bibitem{FW98}
Fryer, C., \& Woosley, S.E., 1998, ApJ, 502, L9.

\bibitem{FWH99}
Fryer, C., Woosley, S.E., \& Hartmann, D.H., 1999, ApJ, 526, 152.

\bibitem{gaisser}
Gaisser, T. K., Halzen, F., Stanev, T., 1995, Phys. Rep., 173.

\bibitem{GG02a}
Guetta D. \& Granot J.,  2003, Phys. Rev. Lett. 90, 201103.
astro-ph/0212045.

\bibitem{GG02b}
Guetta D. \& Granot J., 2003, MNRAS 340, 115, astro-ph/0208156.

\bibitem{GG02c}
Guetta D. \& Granot J., 2003, ApJ 585, 885, astro-ph/0209578.

\bibitem{GSW2}
Guetta D., Spada M., \& Waxman E., 2001a, ApJ 559, 101.

\bibitem{GSW1}
Guetta D., Spada M., \& Waxman E., 2001b, ApJ 557, 339.

\bibitem{Halzen:1999xc}
Halzen, F. \& Hooper, D. W., 1999, ApJ 527, L93. 

\bibitem{Halzen:2002xt}
Halzen, F. \& Hooper, D. W., 2002, Rept.\ Prog.\ Phys.\  65, 1025.

\bibitem{Halzen:1998be}
Halzen, F. \& Saltzberg, D., 1998, PRL 81, 4305. 

\bibitem{xray1}
Heise, J.,  Zand, J. J., Kippen, M. \& Woods, P., 
Gamma-Ray Bursts in the Afterglow Era, Proceedings of the 
International workshop held in Rome, CNR
headquarters, 17-20 October, 2000. 
Edited by Enrico Costa, Filippo Frontera, and Jens Hjorth. Berlin
Heidelberg: Springer, 2001, p. 16. astro-ph/0111246.

\bibitem{IGP02}
Inoue, S., Guetta, D., \& Pacini, F., 2003, 
ApJ, 583, 379.

\bibitem{xray2}
Kippen, R. M. et al.,  2002, astro-ph/0203114.

\bibitem{Kobayashi:2001ve}
Kobayashi, S., Ryde, F., \& MacFadyen, A., 2002
ApJ, 577, 302.

\bibitem{KG}
K\"onigl, A., \& Granot, J. 2002, ApJ, 574, 134. 

\bibitem{Lazzati01}
Lazzati, D. et al., 2001, ApJ 556, 471.

\bibitem{Learned:sw}
Learned, J. G. \& Mannheim, K., 2000,
Ann.\ Rev.\ Nucl.\ Part.\ Sci.\  50, 679.

\bibitem{Lloyd-Ronning:2002jn}
Lloyd-Ronning, N., \& Ramirez-Ruiz, E., 2002,
ApJ, 576, 101.

\bibitem{short1a}
McBreen, S., Quilligan, F., McBreen, B., Hanlon L. \& Watson D.,
astro-ph/0206294.

\bibitem{NPP92}
Narayan, R., Pacz\'nski, B., \& Piran, T. 1992, ApJ, 395, L83.

\bibitem{Norris}
Norris, J. P., Marani, G. F. \& Bonnell, J. T.,2000, ApJ, 534, 248.

\bibitem{short2}
Paciesas, W. S., Preece, R. D. Briggs, M. S. \& Mallozzi, R. S.,
Gamma-Ray Bursts in the Afterglow Era, Proceedings of the 
International workshop held in Rome, CNR
headquarters, 17-20 October, 2000. 
Edited by Enrico Costa, Filippo Frontera, and Jens Hjorth. Berlin
Heidelberg: Springer, 2001, p. 13. astro-ph/0109053.

\bibitem{P98}
Pacz\'nski, B. 1998, ApJ, 494, 45.

\bibitem{Piro00}
Piro, L., et al., 2000, Science 290, 955. 

\bibitem{razzaque}
Razzaque, S., M\'esz\'aros, P. \& Waxman, E., 2002, submitted to PRL, astro-ph/0212536.

\bibitem{RM94}
Rees, M. J., \& M\'esz\'aros, P., 1994, ApJ, 430, L93

\bibitem{Reichart}
Reichart, D. E. et al., 2001, ApJ, 552, 57.

\bibitem{vl3}
Reichart D. E. \&  Lamb, D. Q., 
Gamma-Ray Bursts in the Afterglow Era, Proceedings of the 
International workshop held in Rome, CNR headquarters, 
17-20 October, 2000. 
Edited by Enrico Costa, Filippo Frontera, and Jens Hjorth. Berlin
Heidelberg: Springer, 2001, p. 233.
astro-ph/0103254.

\bibitem{SP97}
Sari, R., \& Piran, T. 1997, ApJ, 485, 270 

\bibitem{Vietri95} 
Vietri M.,  1995, ApJ, 453, 883 

\bibitem{Vietri98a} 
Vietri, M., 1998a, ApJ, 507, 40. 

\bibitem{Vietri98b} 
Vietri, M., 1998b, PRL, 80, 3690. 

\bibitem{VS98}
Vietri, M. \& Stella, L. 1998, ApJ, 507, L45.

\bibitem{Vietri01} 
Vietri M., et al., 2001, ApJ 550, L43.

\bibitem{Vietri03}
Vietri, M., De Marco, D. \& Guetta, D., 2003, 
 ApJ in press, astro-ph/0302144

\bibitem{Waxman95} 
Waxman, E., 1995, PRL, 75, 386.

\bibitem{Waxman97} 
Waxman, E., \& Bahcall, J., 1997, PRL, 78, 2292. 

\bibitem{Waxman00} 
Waxman, E., \& Bahcall, J., 2000, ApJ, 541, 707.

\bibitem{Waxman0}
Gamma-Ray Bursts in the Afterglow Era, Proceedings of the 
International workshop held in Rome, CNR headquarters, 17-20 October, 2000. 
Edited by Enrico Costa, Filippo Frontera, and Jens Hjorth. 
Berlin Heidelberg: Springer, 2001, p. 263.

\bibitem{W93}
Woosley, S.E. 1993, ApJ, 405, 273

\bibitem{Zhang:2002jt}
Zhang, B., \& Meszaros, P., 2002,
ApJ, 581, 1236.




\end{thebibliography}
\end{document}